\documentclass[journal=jctcce,manuscript=article]{achemso}
\setkeys{acs}{articletitle=true}
\usepackage{graphicx}
\usepackage{amsmath}
\usepackage{amsfonts}
\usepackage{amssymb}
\usepackage{amsthm}
\usepackage{braket}
\usepackage{multirow}
\usepackage{xcolor}
\usepackage{achemso}
\usepackage[labelformat=empty]{subfig}
\usepackage{bm}
\usepackage{booktabs}
\usepackage{comment}
\usepackage{mathtools}
\usepackage{upgreek}

\usepackage[utf8]{inputenc}
\newlength\tindent
\DeclareMathOperator{\tr}{Tr}

\newcommand{\bh}[2]{\mathbf{h}^{#1}\mathbf{D}^{#2}}
\newcommand{\bG}[2]{\mathbf{D}^{#1}\mathbf{G}(\mathbf{D}^{#2})}

\graphicspath{{img/}} 

\author{Tommaso Giovannini}
\affiliation{Department of Chemistry, Norwegian University of Science and Technology, 7491 Trondheim, Norway}
\email{tommaso.giovannini@ntnu.no}     
\author{Henrik Koch}
\affiliation{Scuola Normale Superiore,
             Piazza dei Cavalieri 7, 56126 Pisa, Italy.}
\email{henrik.koch@sns.it}

\title[]
  {Energy-Based Molecular Orbital Localization in a Specific Spatial Region}

\begin{document}

\begin{abstract}
We present a novel energy-based localization procedure able to localize molecular orbitals into specific spatial regions. The method is applied to several cases including both conjugated and non-conjugated systems. The obtained localized molecular orbitals are used in a multiscale framework based on the multilevel Hartree-Fock approach. An almost perfect agreement with reference values is achieved for both ground state properties, such as dipole moments, and local excitation energies calculated at the coupled cluster level. The proposed approach is useful to extend the application range of high level electron correlation methods. In fact, the reduced number of molecular orbitals can lead to a large reduction in the computational cost of correlated calculations.
\end{abstract}
\newpage
\begin{center}
\Large \textbf{TOC graphics}
\end{center}
\begin{figure}
\centering
\includegraphics[scale=1]{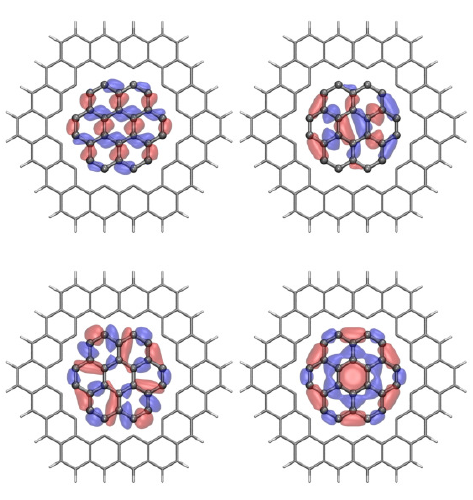}
\end{figure}




\section{Introduction}

Many processes in chemistry take place in a specific spatial region of a molecular system. To rationalize local phenomena, the concept of local occupied molecular orbitals (LMOs) is particularly useful in bridging chemical intuition and theoretical chemistry.\cite{hoyvik2016characterization} The LMOs are very convenient in describing electron correlation, as they can potentially reduce the computational cost of many-body methods.\cite{ma2018explicitly} Among the large variety of different localization procedures developed in the past,\cite{edmiston1963localized,boughton1993comparison,khaliullin2008analysis,khaliullin2007unravelling,aquilante2006fast,hoyvik2012orbital,jansik2011local,hoyvik2015perspective,hoyvik2012trust,ziolkowski2009maximum,gianinetti1996modification,stoll1980use} only a few are able to localize MOs into a specific spatial region of a molecular system.\cite{li2016cluster,hoyvik2016characterization,stoll1980use,
gianinetti1996modification,khaliullin2007unravelling} 

However, such a localization procedure is important when dealing with phenomena taking place in a limited spatial location, for instance local electronic excitations.\cite{li2016cluster,azarias2017modeling} This allows a fragmentation of the target moiety in (at least) two different parts: the active, where the phenomenon takes place, and the inactive, that indirectly influences the active part. Such a partitioning defines the so-called focused models.\cite{mennucci2019multiscale} Most focused models are formulated in terms of quantum mechanical (QM)/classical approaches in which the active-inactive interaction is usually limited to electrostatics.\cite{warshel1976theoretical,cappelli2016integrated,Mennucci12_386,giovannini2019fqfmu} In addition, several fragmentation approaches have been proposed in the last years, demonstrating their capability to treat large molecular systems.\cite{gordon2012fragmentation,collins2015energy,pruitt2014efficient,collins2014combined,pruitt2012fragment} However, all the developed methods are based on approximations, for instance, how fragments are defined, where covalent bonds are cut, how the vacant cap of each fragment is fixed. Also, they are commonly limited to the ground state energy, although some of them have been extended to treat excitation energies,\cite{ding2017embedded,wen2019absolutely,bennie2017pushing} and the interaction between the monomers is usually treated at the electrostatic level.\cite{collins2015energy,chen2020fragment}

In this paper, we are proposing a novel and rigorous method which can provide localized occupied MOs in a specific spatial region of the system. This is achieved by formulating a localization criterion entirely based on energetics. Therefore, our approach differs conceptually from common MO localization procedures,\cite{edmiston1963localized,boughton1993comparison,hoyvik2016characterization,boys1960construction} and from projection-based approaches.\cite{chulhai2017improved,chulhai2018projection,wen2019absolutely,sayfutyarova2017automated} To the best of our knowledge, this is the first time that an energy-based variational method is used to localize MOs in a specific region of the molecule. This also ensures the continuity of the potential energy surface (PES), which is not guaranteed for common localization procedure such as Boys. 

The method is based on multilevel Hartree-Fock (MLHF) theory,\cite{saether2017density,hoyvik2020convergence} which is a rigorous method to partition a molecular system into two different fragments A (active) and B (inactive). Such a partitioning is performed by selecting the number of electrons belonging to the active part, and consequently the number of occupied orbitals is set. As previously reported in Ref. \citenum{saether2017density}, MLHF method differs from most projection-based approaches, either developed in Density Functional Theory (DFT) or in Hartree-Fock (HF) frameworks. The starting point of projection-based approaches is commonly a Self Consistent Field (SCF) calculation on the entire system, and the optimized MOs are then assigned to active and inactive parts by using an priori orbital assignment.\cite{chulhai2017improved,chulhai2018projection,wen2019absolutely,culpitt2017communication,hegely2016exact,huzinaga1971theory,goodpaster2010exact,goodpaster2012density,goodpaster2014accurate,manby2012simple,sun2016quantum} The quantum embedding Hamiltonian is then constructed by including an exact or approximated embedding operator.\cite{huzinaga1971theory,goodpaster2010exact,goodpaster2012density,goodpaster2014accurate,manby2012simple,sun2016quantum} In particular, we want to highlight that the selection of the active MOs is usually performed by using a predefined threshold metric, which however may cause the a wrong MO selection, as reported by Kallay and co-workers.\cite{hegely2016exact} 

In MLHF, no approximations are introduced neither in the bond fragmentation or in the interaction energy between the active and inactive parts. In this paper, we show that our novel energy-based localization procedure can provided MOs that are well localized on the specified active fragment, without carring out an a priori orbital assignment. We demonstrate that our approach can provide not only accurate ground state properties, but also accurate local excitation energies (calculated at the coupled cluster (CC) level). Within this scheme, coupled cluster ground and excited state calculations are performed using the MOs of the active fragment only, thus the intrinsic computational cost of high level calculations is reduced, similarly to other multilevel methods.\cite{myhre2016multilevel,folkestad2019multilevel} For the same reason, the accuracy of the computed local excitations depends crucially on the quality of the LMOs.

The manuscript is organized as follows. In the next section, the MLHF theory is briefly summarized and the energy-based orbital localization is discussed. Then, the computational procedure and the numerical applications are presented, with particular emphasis on the spread of the obtained localized MOs and on the accuracy of the novel approach in predicting local properties such as dipole moments and excitation energies. Summary and conclusions of the present work end the manuscript.

\section{Theory}

The active-inactive partitioning in MLHF is realized by decomposing the density of the whole system into active and inactive densities ($\mathbf{D} = \mathbf{D}^A + \mathbf{D}^B$). The total HF energy can be written as:
\begin{align}
E & = \tr \textbf{hD} + \frac{1}{2}\tr \textbf{D} \textbf{G}(\textbf{D}) +h_{nuc} \nonumber \\
  & = \tr \textbf{hD}^A + \frac{1}{2}\tr \textbf{D}^A \textbf{G}(\textbf{D}^A) + \tr \textbf{D}^A \textbf{G}(\textbf{D}^B) + \tr \textbf{hD}^B + \frac{1}{2}\tr \textbf{D}^B \textbf{G}(\textbf{D}^B) + h_{nuc} ,
\label{eq:mlhf_tot}
\end{align}
where $\mathbf{h}$ and $\mathbf{G}$ are the usual one and two-electron matrices, and $h_{nuc}$ is the nuclear repulsion. The $\mathbf{G}(\mathbf{D})^X$ with $X = \{A,B\}$ matrix is defined as:

\begin{align}
G_{\mu\nu}(\mathbf{D}^X) & = \sum_{\sigma\tau} D^X_{\sigma\tau} \left ( (\mu\nu | \sigma\tau) - \dfrac{1}{2} ( \mu\tau | \sigma\nu ) \right ).
\end{align}
The main idea of MLHF is to optimize the density of fragment $A$ in the field generated by the density $B$, which is kept fixed. This procedure is performed by minimizing the energy (see Eq. \ref{eq:mlhf_tot}) in the MO basis of the active part, reducing the dimensionality of the problem.
In MLHF, the Fock matrix in AO basis is expressed by differentiating Eq. \ref{eq:mlhf_tot} with respect to $\mathbf{D}^A$:

\begin{equation}
\label{eq:mlhf_fock}
F_{\mu\nu} = h_{\mu\nu} + G_{\mu\nu}(\mathbf{D}^A) + G_{\mu\nu}(\mathbf{D}^B) \ .
\end{equation} 

In Eq. \ref{eq:mlhf_fock}, the last term $G_{\mu\nu}(\mathbf{D}^B)$ is a one-electron contribution, because the $\mathbf{D}^B$ density is kept frozen during the SCF procedure.

Equation \ref{eq:mlhf_tot} is formally equal to the full HF energy when $\mathbf{D}$ is the converged SCF density for the entire system. However, Eq. \ref{eq:mlhf_tot} does not have an apparent physical interpretation because the different energy terms are not assigned to the individual fragments. Such a physical insight can be achieved by dividing the one-electron term into the kinetic ($\mathbf{T}$) and the electron-nuclear attraction operators for the two parts ($\mathbf{V}^A$ and $\mathbf{V}^B$).  Thus, Eq. \ref{eq:mlhf_tot} can be written as:
\begin{align}
E & = \underbrace{\tr \mathbf{h}^A\mathbf{D}^A + \dfrac{1}{2}\tr \mathbf{D}^A\mathbf{G}(\mathbf{D}^A) + h^{A}_{nuc}}_{E_A} + \underbrace{\tr \mathbf{h}^B\mathbf{D}^B + \dfrac{1}{2}\tr \mathbf{D}^B\mathbf{G}(\mathbf{D}^B) + h^{B}_{nuc}}_{E_B} + \nonumber \\
  & + \underbrace{\tr \mathbf{V}^B\mathbf{D}^A + \tr \mathbf{V}^A\mathbf{D}^B + \tr \mathbf{D}^A \mathbf{G} (\mathbf{D}^B) + h^{AB}_{nuc}}_{E_{AB}} \ ,  
 \label{eq:mlhf_energies_ab}
\end{align}
where $h^{A}_{nuc}$, $h^{B}_{nuc}$ and $h^{AB}_{nuc}$ are nuclear repulsion terms; $E_A$ and $E_B$ are the energies of the two fragments, whereas $E_{AB}$ is the interaction energy. The latter term is composed of the electron-nuclear attraction between $A$ and $B$ and viceversa, and the coulomb and exchange interactions between the two fragments. 

Although Eq. \ref{eq:mlhf_energies_ab} is equivalent to Eq. \ref{eq:mlhf_tot}, it permits the definition of our localization procedure in a fragment-based model such as MLHF. In fact, an additional SCF procedure can be performed to optimize the energy of part A and/or B, in the occupied space of both fragments, i.e. without changing the total energy. The procedure is general and can be performed on the basis of any density matrix $\mathbf{D}$ that is decomposed into two densities belonging to two fragments. 

Two alternatives can be defined. First, the energy term $E_A$ (see Eq. \ref{eq:mlhf_energies_ab}) can be minimized (denoted MLHF-A). In such a case, the Fock matrix reads:

\begin{equation}
F_{\mu\nu} = h^A_{\mu\nu} + G(\mathbf{D}^A)_{\mu\nu} \ .
\end{equation}

In the second approach (called MLHF-AB), the sum of $E_A$ and $E_B$ is minimized. From the computational point of view, $E_A+E_B$ can be rewritten by considering that the total density $\mathbf{D} = \mathbf{D}^A + \mathbf{D}^B$ remains constant during occupied-occupied rotations. This means that $\mathbf{D}^B$ can be expressed in terms of it as $\mathbf{D}^B = \mathbf{D} - \mathbf{D}^A$. Therefore, the sum of $A$ and $B$ energies reads: 

\begin{align}
E_A + E_B & = \tr \bh{A}{A} + \dfrac{1}{2}\tr \bG{A}{A} + \tr \bh{B}{} - \tr \bh{B}{A} + \nonumber \\
 & +\dfrac{1}{2}\tr \bG{}{} + \dfrac{1}{2}\tr \bG{A}{A} - \tr \bG{A}{} = \nonumber \\
 & = \tr (\mathbf{V}^A - \mathbf{V}^B) \mathbf{D}^A + \tr \bG{A}{A} - \tr \bG{A}{} + \tr \bh{B}{} + \dfrac{1}{2} \tr \bG{}{} \,
\label{eq:MLHF-Ab}
\end{align} 

where the last two terms depend on the total density only, and are therefore constant energy terms. The first three terms are instead similar to the MLHF energy contributions (see Eq. \ref{eq:mlhf_tot}), because they are characterized by one-electron and two-electron contributions involving the active density only. The third term $\tr \bG{A}{}$ is instead the two-electron interaction between the active and the constant total density $\mathbf{D}$. The Fock matrix of the active part can be written as:

\begin{align}
F_{\mu\nu} & = V^A_{\mu\nu} - V^B_{\mu\nu} + G_{\mu\nu}(\mathbf{D}^A) - G_{\mu\nu}(\mathbf{D}^B) \nonumber \\
& =V^A_{\mu\nu} - V^B_{\mu\nu} + 2G_{\mu\nu}(\mathbf{D}^A) - G_{\mu\nu}(\mathbf{D}) \ ,
\end{align}

which is again in the same form as Eq. \ref{eq:mlhf_fock}, because it is characterized by one-electron contributions ($V_{\mu\nu}^A - V_{\mu\nu}^B$), a two-electron contribution on the active density $2G_{\mu\nu}(\mathbf{D}^A)$, and a constant contribution due to the total density $G_{\mu\nu}(\mathbf{D})$.

Notice that minimizing the sum of A and B parts in MLHF-AB (see Eq. \ref{eq:MLHF-Ab}) is equivalent to maximizing the interaction energy $E_{AB}$. Physically, this means that the repulsion between the two parts is maximized, and the occupied orbitals obtained by this scheme are those maximally located in the two fragments.

\section{Computational Procedure}

The two approaches are implemented in a development version of the electronic structure program $e^\mathcal{T}$,\cite{eT_arxiv} and follow the computation protocol graphically depicted in Fig. \ref{fig:protocol}:

\begin{figure}[htbp!]
\centering
\includegraphics[width=.4\textwidth]{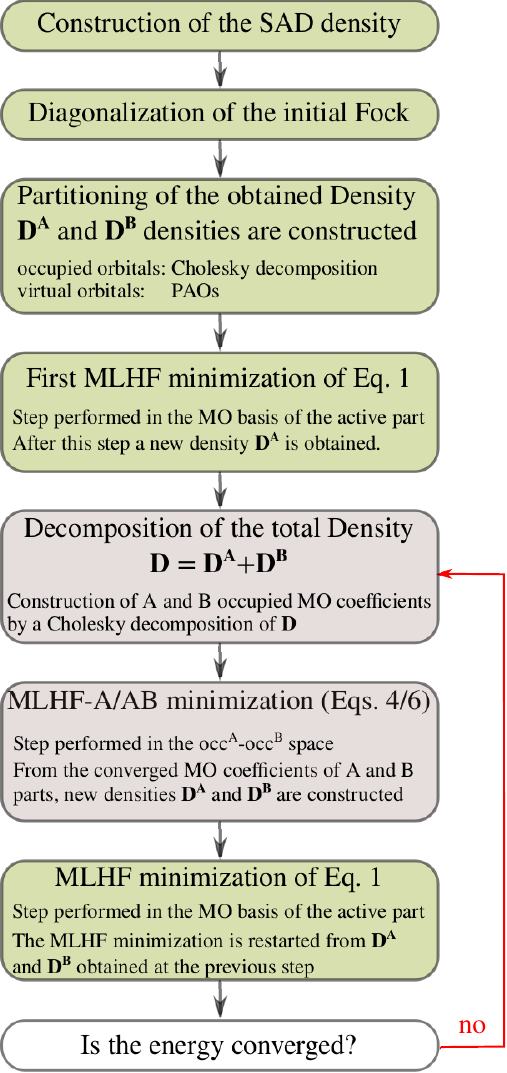}
\caption{Graphical view of the computational procedure.}
\label{fig:protocol}
\end{figure}

\begin{enumerate}
\item Construction of the initial density by means of superposition of atomic densities (SAD), followed by a diagonalization of the initial Fock matrix.
\item Partitioning of the resulting density into $A$ and $B$ densities, using Cholesky decomposition for the active occupied orbitals and projected atomic orbitals (PAOs) for active virtual orbitals.\cite{aquilante2011cholesky,sanchez2010cholesky,koch2003reduced,christiansen2006coupled,hoyvik2015perspective} 
We note that the Cholesky decomposition of the total density $\mathbf{D}$ into $\mathbf{D}^A$ and $\mathbf{D}^B$ is a mathematical method to decompose a matrix, which is unique if the same pivots are used. In this work, the Cholesky decomposition is performed by selecting the diagonals corresponding to the basis functions which are centered on the active atoms.\cite{saether2017density,sanchez2010cholesky} In particular, $\mathbf{D}^A$ is written in the AO basis $\{\alpha,\beta\}$ as:\cite{myhre2014multi}

\begin{align}
D^{A}_{\alpha\beta} & = \sum_{IJ} D_{\alpha I} \tilde{D}^{-1}_{IJ} D_{\beta J}  \nonumber \\
& = \sum_{I} L_{\alpha I} L_{\beta I}\label{eq:cholesky}
\end{align}

where $I$ and $J$ are the diagonal elements which are decomposed,  $\tilde{\mathbf{D}}$ is the submatrix of $\mathbf{D}$ containing the diagonal elements, and $L_{\alpha I}$ are the Cholesky orbitals. The Cholesky threshold is chosen so that the number of $\mathbf{D}$ diagonal elements corresponds to the correct number of occupied orbitals of the active fragment. As a result of the decomposition, the active Cholesky MOs are obtained and the active density matrix $\mathbf{D}^A$ is trivially constructed (see Eq. \ref{eq:cholesky}). The inactive density $\mathbf{D}^B$ is instead obtained as a difference between the total density $\mathbf{D}$ and the active one $\mathbf{D}^A$.
\item The energy defined in Eq. \ref{eq:mlhf_tot} is minimized in the MO basis of the active part. 
\item The total density $\mathbf{D}$ is constructed by summing the MLHF converged density $\mathbf{D}^A$ and the inactive density $\mathbf{D}^B$, and active/inactive occupied orbitals are obtained by a Cholesky decomposition. Again, the total density $\mathbf{D}$ is Cholesky decomposed by considering the diagonals belonging to the active atoms. The inactive density $\mathbf{D}^B = \mathbf{D} - \mathbf{D}^A$ is then Cholesky decomposed by considering the diagonals belonging to the inactive atoms. From the two Cholesky decompositions, active and inactive MOs are obtained and the occupied-occupied space is defined.
\item The energy of A (in MLHF-A) and B (in MLHF-AB) are minimized (Eq. \ref{eq:mlhf_energies_ab}) in the MO space defined by the occupied orbitals of the active and inactive parts in an SCF procedure.
\item From MLHF-A/AB occupied MO coefficients, active and inactive densities are constructed and a new MLHF calculation is restarted from point 3, until convergence is reached. For all the results reported in this paper, three macrocycles MLHF -- MLHF-A/AB are sufficient to reach full convergence of the energy. It is worth noticing that since the MLHF calculation is restarted from the MO coefficients obtained at the 5-th step, the total computational cost of MLHF-A/AB is only twice than a standard MLHF calculation.
\end{enumerate}

\section{Numerical Applications}

\begin{figure}[htbp!]
\centering
\includegraphics[width=1\textwidth]{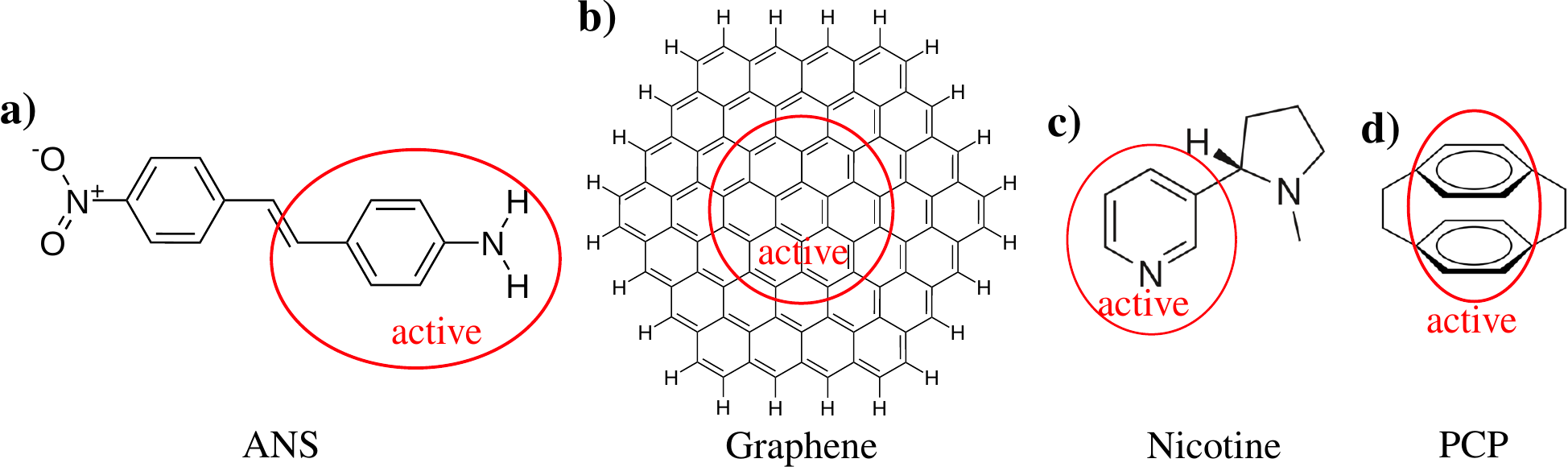}
\caption{Molecular structures of ANS (\textbf{a}), graphene (\textbf{b}), nicotine (\textbf{c}), and PCP (\textbf{d}). The active parts used in MLHF calculations are highlighted.}
\label{fig:structures}
\end{figure}

The capabilities of MLHF-A/AB are illustrated for four different molecular moieties, that have previously been studied theoretically and experimentally.\cite{egidi2014benchmark,marder1991approaches,grigorenko2012graphene,gibson20032,gleiter2006modern,grimme2004importance,demissie2016pcp,bachrach2011pcp} Those are 4-amino-4'-nitrostilbene (ANS), a part of a graphene sheet (which is indeed a graphene quantum dot), (S)-nicotine (in its most stable conformer\cite{egidi2014benchmark}), and [2,2]paracyclophane (PCP) (see Fig. \ref{fig:structures} for the molecular structures). Molecular geometries of ANS and nicotine are optimized at the B3LYP/aug-cc-pVDZ by using Gaussian16 package.\cite{gaussian16} The graphene sheet is constructed by setting the C-C distance to 1.42 \AA, and the C-H distance to 1.07 \AA.\cite{neto2009electronic} The PCP geometry is taken from Ref. \citenum{demissie2016pcp}. Graphene and ANS are chosen because they are conjugated systems. The conjugation is broken by our definition of the active regions as depicted in Fig. \ref{fig:structures}a-b (in both cases the bonding electrons are assigned to the inactive part). In case of nicotine and PCP single covalent bonds are cut and the bonding electrons are assigned to the active fragment (see Fig. \ref{fig:structures}c-d). Hereby we demonstrate the generality of our procedure, which can be applied to different cases (single/double bond cutting) and to different definitions of the active region.

Nicotine and ANS calculations are performed by combining MLHF(/CC2) with aug-cc-pVDZ basis sets. The active part of the graphene sheet is described using cc-pVTZ basis set, whereas its inactive part with cc-pVDZ basis set. The PCP MLHF/CC2 calculations are performed with the triple-zeta quality 6-311G(d,p) basis set.\cite{bachrach2011pcp} Notice that the different basis sets are chosen so to demonstrate the reliability of our approach in combination with diffuse/polarization functions. 

The orbital second central moment (orbital variance) is used to quantitatively characterize orbital locality. The second central moment $\mu^p_2$ of an MO $\varphi_p$ is defined as:\cite{boys1960construction}

\begin{equation}
\mu^p_2 = \braket{\varphi_p|\mathbf{r}^2|\varphi_p} - \braket{\varphi_p|\mathbf{r}|\varphi_p}^2 \ .
\label{eq:centralmoment}
\end{equation}

The orbital spread $\sigma_p$ is defined as the square root of $\mu^p_2$. We also defined $\xi$ as the average value of $\sigma_p$, i.e. $\xi$ is a measure of the mean locality of the considered set of MOs.\cite{hoyvik2016characterization} In this paper, MLHF-A/AB MOs are compared with canonical MLHF ones (named Cholesky because they are obtained through a Cholesky decomposition of the initial density matrix), that are also localized with the Boys procedure (Cholesky-Boys).\cite{boys1960construction} Notice that in Boys localization, the sum over $p$ of $\mu^p_2$ in Eq. \ref{eq:centralmoment} is minimized,\cite{boys1960construction} and the obtained MOs can therefore be used as reference for both MLHF-A and MLHF-AB approaches. 

\subsection{MLHF-A/AB Localized MOs}

\begin{figure}[htbp!]
\includegraphics[width=1\textwidth]{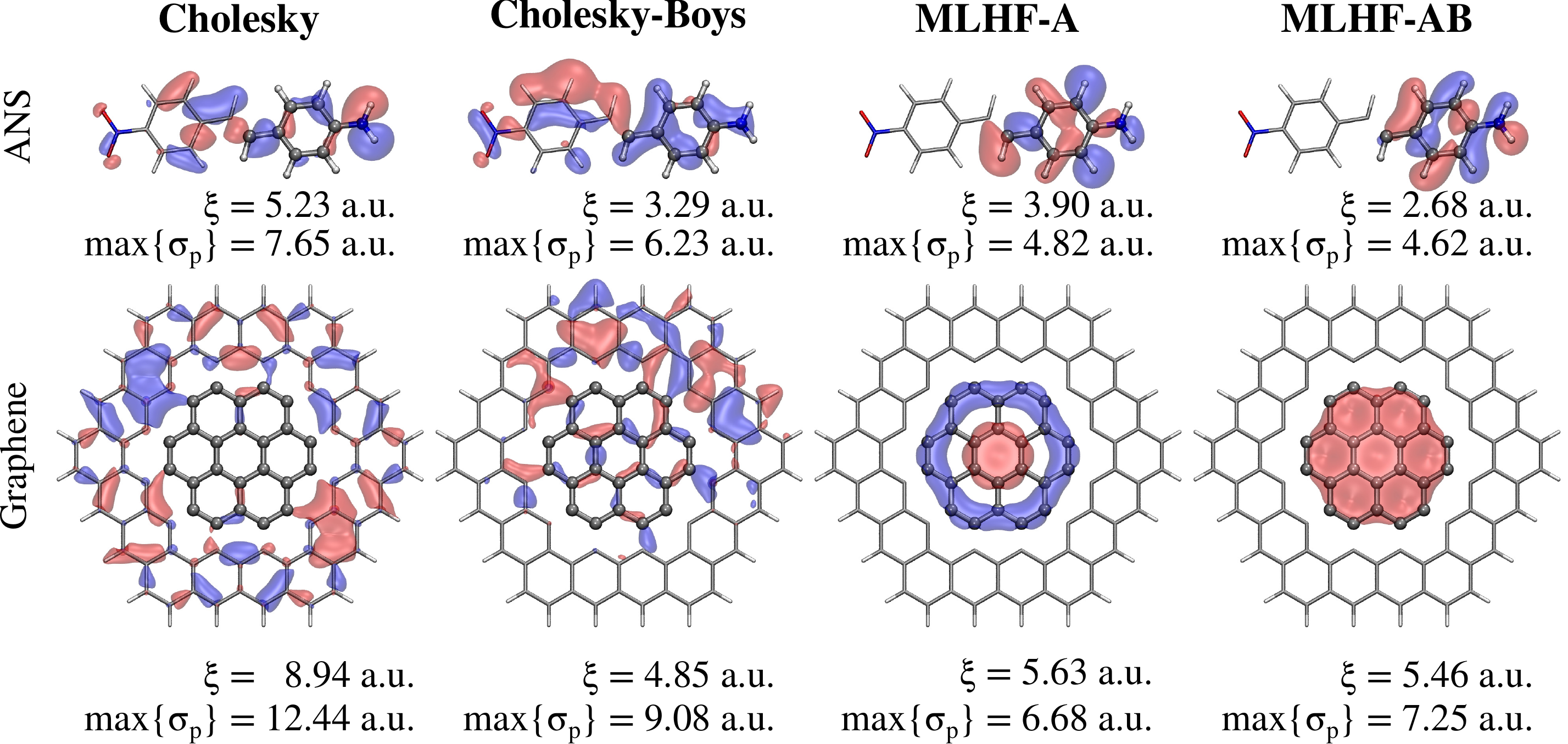}
\caption{Graphical depiction of the most delocalized MOs of ANS (top) and graphene (bottom) calculated by MLHF (Cholesky), Cholesky-Boys, MLHF-A and MLHF-AB methods. Computed $\xi$ and maximum MO spread for all methods are also given.}
\label{fig:mos}
\end{figure}

The most delocalized MOs of ANS and graphene are depicted in Fig. \ref{fig:mos}, and the value of $\xi$ and the maximum $\sigma_p$ are also reported (see Sec. S1.1 and S1.2 in Supporting Information - SI for the spreads of all occupied valence orbitals). First, we notice that in both ANS and graphene, Cholesky orbitals have the largest spread on average ($\xi$) and the largest maximum MO spread (max$\{\sigma_p\}$). As expected, both parameters are reduced by Cholesky-Boys. The MOs calculated by both methods are delocalized over the whole molecule (for both ANS and Graphene).

\begin{figure}[htbp!]
\includegraphics[width=.82\textwidth]{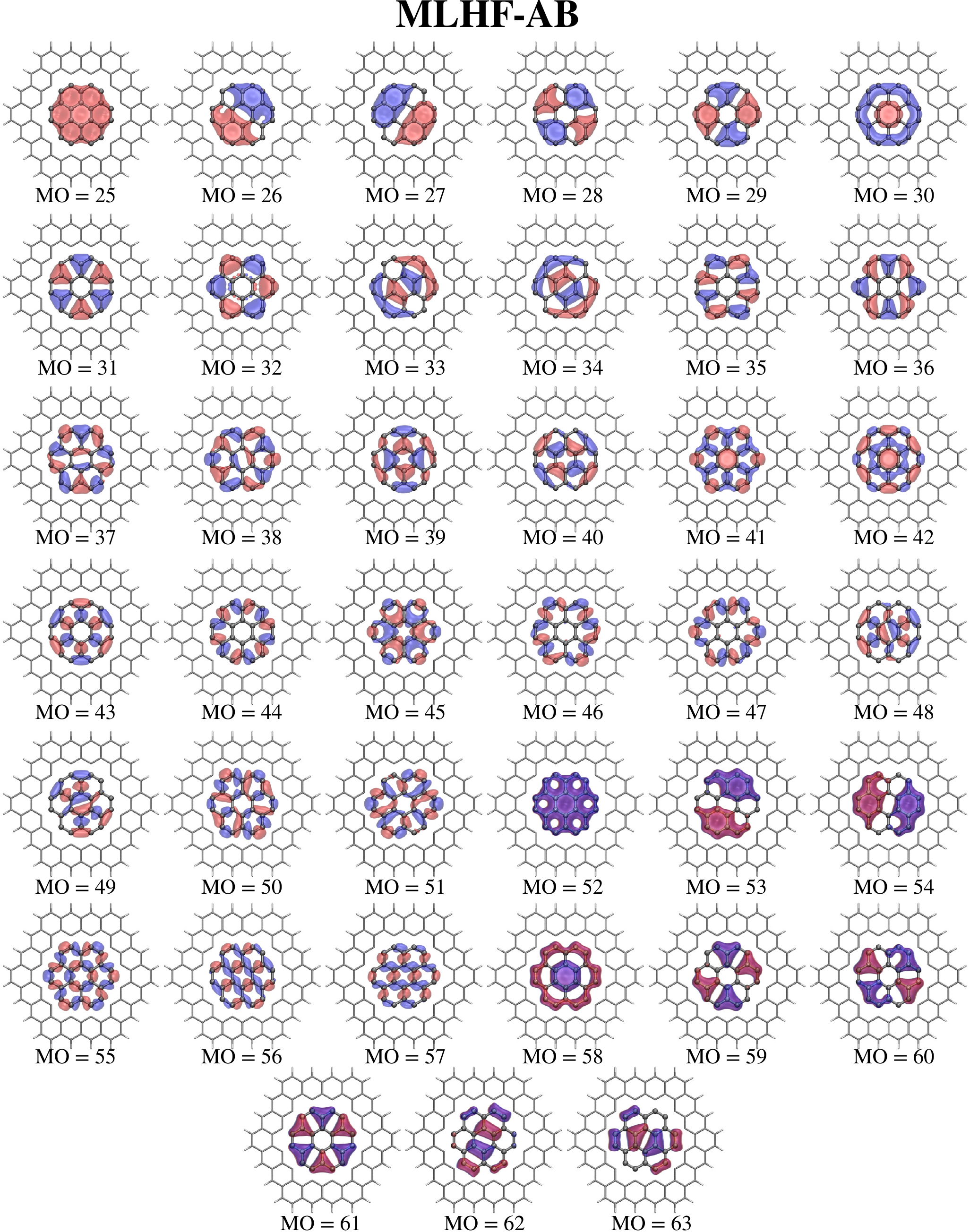}
\caption{Graphical depiction of the active MOs of graphene as predicted by MLHF-AB.}
\label{fig:graph_mos}
\end{figure}

A completely different picture arises when MLHF-A/AB methods are employed. The MOs obtained by both the latter approaches are well-localized on the active part only, and the values of $\xi$ and max$\{\sigma_p\}$ are reduced compared to the corresponding Cholesky counterparts. It is also worth noticing significant differences between MLHF-A and MLHF-AB in particular for ANS. In fact, the most diffuse MLHF-A MO has a tail connecting active and inactive fragments, which should be absent since bonding electrons are assigned to the inactive part. Such a tail is completely absent in the case of MLHF-AB. From a physical point of view, this is not surprising. In fact, in the MLHF-AB procedure (see Eq. \ref{eq:mlhf_energies_ab}) the occupied orbitals of the active and inactive fragments are rotated in order to minimize the sum of the two energies. As stated above, such a rotation corresponds to maximizing the interaction energy between the two parts, i.e. to maximizing the repulsion between them. As a consequence, the active occupied orbitals calculated by MLHF-AB are more localized on the active part. 

The same does not apply to MLHF-A where the active energy is minimized in the occupied-occupied space. Thus no constraints are imposed neither on the inactive energy or on the interaction energy. However, notice that a few MOs have tails in both methods (see Sec. S1.1 and S1.2 in SI), since MLHF-A/AB orbitals are orthogonal.\cite{hoyvik2016characterization} The tails can be reduced by further localizing MLHF-A/AB orbitals using standard localization procedures. Notice also that the MLHF-AB $\xi$ and max$\{\sigma_p\}$ for ANS computed by using cc-pVTZ and aug-cc-pVDZ give very similar results, thus showing the consistency of our approach when diffuse functions are included (see Table S1 in the SI). The observations for ANS and graphene also apply to nicotine and PCP, whose MOs and corresponding spreads are reported in Sec. S1.3 and S1.4 in SI. For the latter systems, the $\xi$ for Cholesky-Boys are lower than the corresponding MLHF-A/AB counterparts, but the MOs also spread in the inactive region. To illustrate the robustness of our approach, a different definition of active/inactive parts of ANS is also investigated (see Sec. S1.1.2 in the SI). The calculated results confirm the findings here discussed. 

In Fig. \ref{fig:graph_mos}, we report the local MOs belonging to the active fragment of graphene as calculated by using the MLHF-AB method. All the MLHF-AB local MOs are well-confined in the active part, and the symmetry of each orbital is evident. 
As a final comment, it is worth noticing that MLHF-AB orbitals may be further confined by using common localization procedures, such as Boys. The resulting local MOs will be more localized on the active atoms, than those obtained by performing an hypothetical localization on the MOs resulting from SCF procedure of the entire system.

\subsection{Ground State Dipole Moments}

\begin{figure}[htbp!]
\includegraphics[width=.60\textwidth]{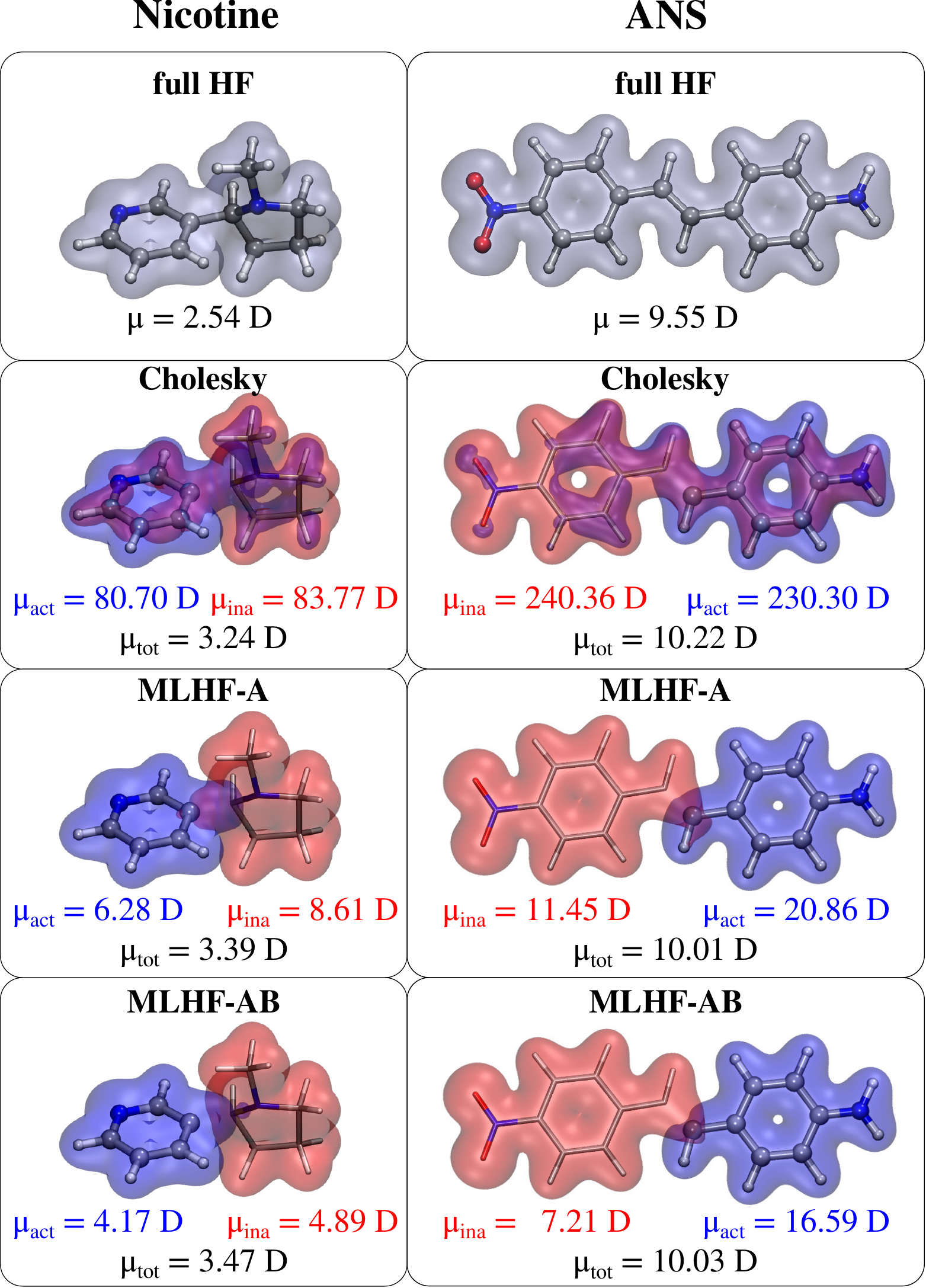}
\caption{Nicotine (left) and ANS (right) MLHF (Cholesky)/MLHF-A/MLHF-AB dipole moments. For MLHF calculations, active (blue) and inactive densities (red) and their corresponding dipole moments are also given. Full HF dipole moment and molecular density are also shown.}
\label{fig:nic_ans_dipoles}
\end{figure}

The MLHF-A and MLHF-AB methods are also applied to calculate ground state properties. We study the dipole moments of the active and inactive regions, together with the total dipole moments predicted by Cholesky, MLHF-A and MLHF-AB. We are not reporting the results using Cholesky-Boys because a rotation among the active occupied orbitals does not change the active density and the density-related properties, such as the dipole moment. The numerical values of active, inactive and total dipole moments for both nicotine and ANS are reported in Fig. \ref{fig:nic_ans_dipoles}, together a graphical representation of active (blue) and inactive (red) densities. Full HF densities and ground state dipole moments for both molecules are also given and used as reference. Dipole moments of both graphene and PCP structures are not reported because they are zero due to symmetry.

The MLHF (Cholesky) predicts large active/inactive dipole moments for both systems; for nicotine, they are almost 80 D, whereas for ANS almost 230 D. Such large dipole moments can be explained by investigating the spatial extension of active and inactive densities, which in both molecules are overlapping. This is due to the fact that the initial Cholesky decomposition defines an inactive density that overlaps with the active part and viceversa. Such issues are solved by MLHF-A/AB. Although both methods start from the same densities as those obtained by MLHF (Cholesky), the occupied-occupied rotations make the active and inactive densities more confined in their specific spatial regions, with a partial overlap limited to the bonding regions. As a consequence, the calculated dipole moments are very similar, in particular for ANS. 

In both molecules, the MLHF-AB dipole moments are much lower than the corresponding ones for MLHF-A (see Fig. \ref{fig:nic_ans_dipoles}). This is due to the maximization of the active-inactive repulsion in MLHF-AB. Thus, a further confinement of the two densities in their specific spatial regions takes place.This can be appreciated by inspecting the bonding regions in both nicotine and ANS (Fig. \ref{fig:nic_ans_dipoles}), showing that the overlap between active and inactive densities is lower in MLHF-AB than in MLHF-A. Notice also that the bonding electrons of nicotine are assigned to the active part. Therefore, the active (blue) density in Fig. \ref{fig:nic_ans_dipoles} defines the bond, whereas for ANS the opposite applies. In case of both nicotine and ANS, the total dipole moments are in very good agreement with the full HF reference value, with the largest discrepancy given by MLHF-AB for nicotine (error = 37\%). The numerical value of the total dipole moments can be improved by using a different initial guess density, as for instance superposition of molecular densities (obtained by means of molecular fractionation with conjugate caps\cite{zhang2003molecular}). 

\subsection{MLHF-AB vs. projection-based approaches}

In this section, the MLHF-AB model is compared to HF projection-based approach (called projected-HF). In particular, we compute ground state energies and dipole moments of the active fragment of nicotine and ANS molecules as a function of the elongation of the active-inactive covalent bond, which is cut by the partitioning into two fragments. The projected-HF results are obtained by first performing a SCF calculation on the entire structure. Then, the SCF MOs are localized by Boys localization procedure, and the MOs are assigned to the active and the inactive part. In order to be comparable with MLHF-AB results, the number of occupied MOs belonging to the active fragment is calculated by setting the number of the electrons in the active part ($n_o = \dfrac{n_{el}}{2}$) similarly to MLHF-AB calculations. The MOs belonging to the active fragment have to be selected on some mathematical criterion. Here, we calculate the percentage ($p^A_i$) of the $i$-th MO in the active part $A$ as:

\begin{equation}
p^A_i = \dfrac{\sum_{\mu \in A} C^2_{i\mu}}{\sum_{\mu \in A,B} C^2_{i\mu}}\cdot 100 \ ,
\end{equation}

where, the $C_{i\mu}$ is the MO coefficient of the $i$-th MO in AO basis $\{\mu\}$. The $n_o$ active MOs in projected-HF calculations are then selected as those having the highest percentage in the active atoms. It is worth pointing out that the active MOs in projected-HF models can also be selected as those having a percentage $\geq$ 50\%, instead of fixing the number of active MOs to $n_o$. However, when applied to PES studies, such a choice leads to unavoidable PES discontinuities because a different number of active MOs may be selected depending on the active-inactive distance. Also, different methods to calculate the MO percentage in A can be arbitrarily chosen, thus the results are not unique. For these reasons, we prefer to keep the number of active MOs fixed to $n_o$. We notice that such an 	arbitrariness is almost absent in MLHF-AB calculations, which only depend on the active-inactive partitioning of the electrons in the studied system. 

In projected-HF method, once the active MO coefficients are selected, the active density is  constructed ($D_{\mu\nu}^A = \sum_{ij} C_{\mu i}C_{j\nu}$), and the active energy is calculated as the $E^A$ term in Eq. \ref{eq:mlhf_energies_ab}.
 
\begin{figure}[htbp!]
\includegraphics[width=.7\textwidth]{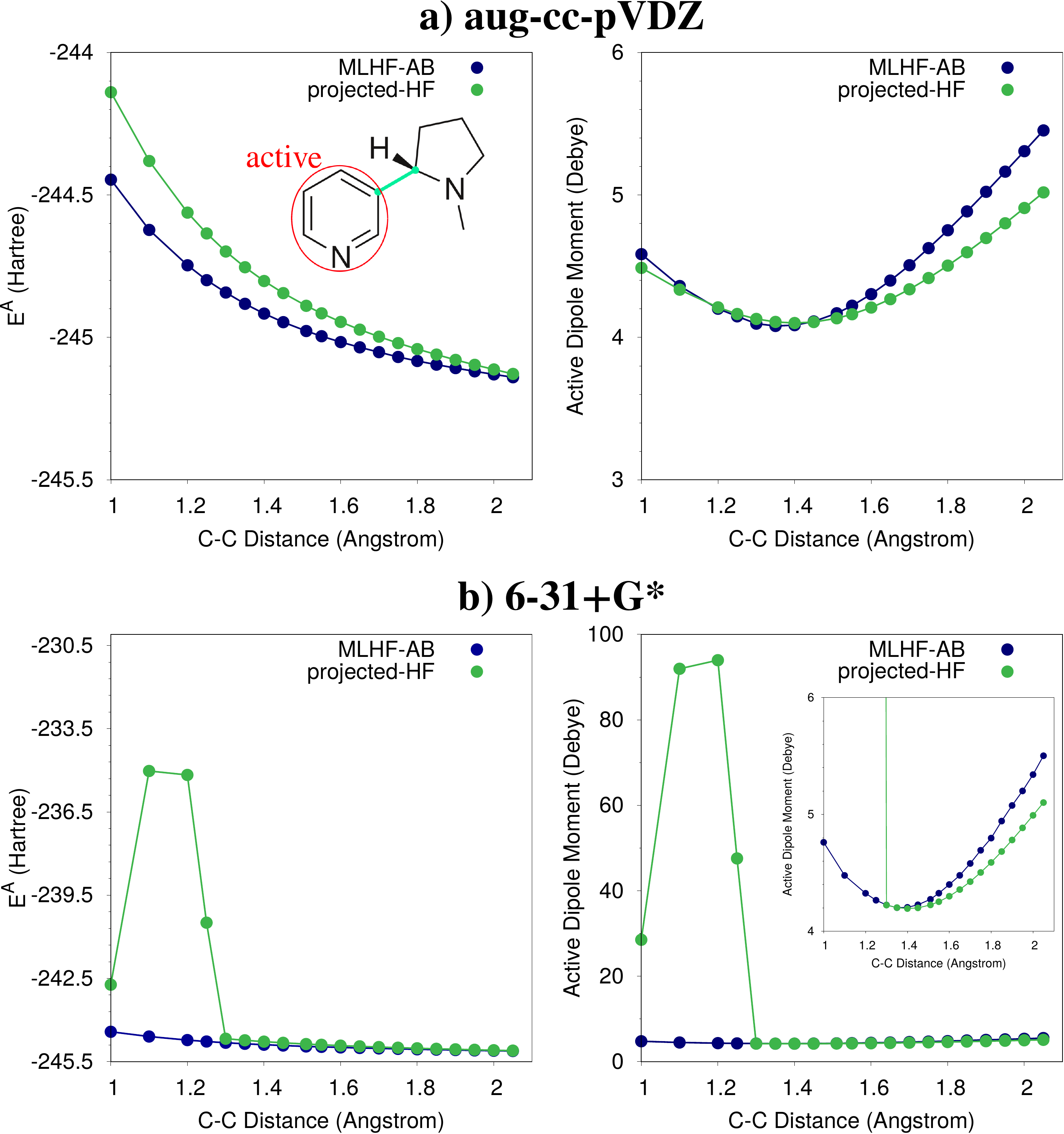}
\caption{MLHF-AB and projected-HF active ground state energy $E^A$ (left) and dipole moment of the active part (right) of nicotine as calculated by using aug-cc-pVDZ (a) or 6-31+G* (b) basis sets.}
\label{fig:proj_mlhfab}
\end{figure}

In Fig. \ref{fig:proj_mlhfab}, MLHF-AB and projected-HF methods are applied to the calculation of ground state energies and dipoles of the active fragment of nicotine as a function of the elongation of the single covalent bond connecting the active and inactive parts. In particular, two different basis sets, aug-cc-pVDZ (panel a) and 6-31+G* (panel b), are used. The equilibrium distance is 1.51 \AA, and the covalent is bond is varied from 1.00 to 2.1 \AA. In both approaches, the number of active occupied MOs $n_o$ is fixed to 21, because, as stated above, the bonding electrons are assigned to the active fragment. 

The results reported in Fig. \ref{fig:proj_mlhfab}a, clearly show that by using the aug-cc-pVDZ basis set both MLHF-AB and projected-HF do not display any PES discontinuities (left panel). Also the energy difference between the two approaches rapidly decreases as the active-inactive distance is elongated. At the equilibrium geometry the MLHF-AB--projected-HF energy difference is of about 0.1 Hartree, with the MLHF-AB energy that is lower than projected-HF one at all the considered distances. This is not surprising and results from the minimization procedure in MLHF-AB (see Eq. \ref{eq:MLHF-Ab}). The dipole moment of the active part is reported in the right panel of Fig. \ref{fig:proj_mlhfab}a. Also in this case, the curves obtained by using both approaches do not display any discontinuities, and a difference of about 0.3 Debye is reported at the equilibrium geometry. 

A different picture arises when the 6-31+G* basis set is used (Fig. \ref{fig:proj_mlhfab}b). In this case, the projected-HF PES clearly displays a large discontinuity at small active-inactive distances, both in the ground state energy (left) and in the dipole moment of the active part (right). Such a discontinuity reflects a discontinuity in the Boys space, which is common in MO localization procedures as it has been reported in different contexts.\cite{russ2004potential} The discontinuity is completely absent in MLHF-AB. However, in the proximity of the equilibrium geometry, both approaches do not display any discontinuities. At the equilibrium geometry, the MLHF-AB energy is lower than the projected-HF by about 0.1 Hartree, and the MLHF-AB--projected-HF active dipole moment difference is about 0.05 Debye. 
The present analysis shows that the MLHF-AB PES is always continuous, whereas the projected-HF PES can display some discontinuities depending on the selected basis set. Notice that the results discussed for nicotine also apply to the case of ANS molecule (see Fig. S27 given as SI). For both nicotine and ANS, the average spread ($\xi$) of the LMOs used in projected-HF is smaller than in MLHF-AB (1.71 a.u. vs. 2.11 a.u. (nicotine) and 1.70 a.u. vs. 2.68 a.u. (ANS)). This is expected as the Boys localization minimizes the orbital spread $\xi$, and therefore provides the LMOs with the lowest $\xi$. Also, the MLHF-AB is not intended to provide the most localized MOs overall, but the most localized MOs in a specific spatial region. In passing, we note that MLHF-AB PES can display discontinuities depending on the initial Cholesky decomposition. However, this can be avoided by selecting the same pivots during the Cholesky decomposition of the initial density. 

\subsection{Coupled Cluster absorption energies}

\begin{figure}[htbp!]
\includegraphics[width=.85\textwidth]{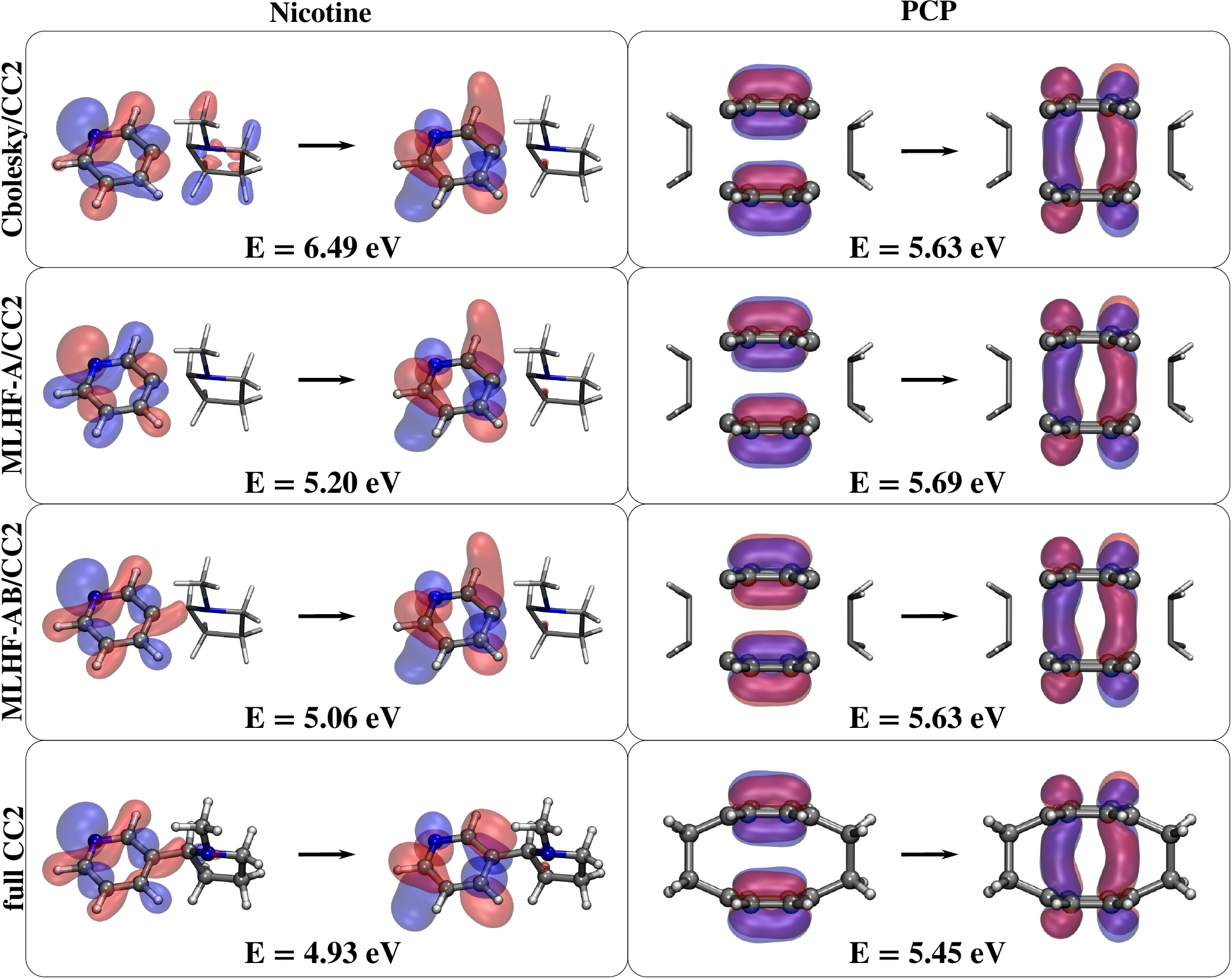}
\caption{Nicotine (left) and PCP (right) MLHF (Cholesky)/MLHF-A/MLHF-AB and full CC2 excitation energies for the depicted transitions.}
\label{fig:excitations}
\end{figure}

As a final application of MLHF-A and MLHF-AB, we select two local transitions, i.e. occurring in the selected active parts, exhibited by nicotine and PCP (for which we investigate a though-space charge transfer excitation,\cite{grimme2004importance} see Fig. \ref{fig:excitations}). 

Local excitations are a perfect test case for demonstrating the capabilities of both approaches proposed here. In fact, the quality of the localized orbitals is crucial for obtaining a reliable excitation energy. In this work, the excitation energies are computed using CC2\cite{christiansen1995second} for the active MOs. 

The MLHF (Cholesky) and MLHF-A/AB results are compared with full CC2 reference excitation energies. From the inspection of Fig. \ref{fig:excitations}, it is clear that MLHF-A and MLHF-AB transition energies are in reasonable agreement with the reference values, in particular in case of MLHF-AB. For nicotine, MLHF (Cholesky) is completely unable to reproduce the full CC2 excitation energy, because of the non-locality of the occupied MO.  For PCP, all investigated methods give similar excitation energies, as the occupied orbitals are similarly reproduced by all approaches. We note a relatively large difference between MLHF-A/AB and full CC2 excitation energies. This discrepancy can be reduced by systematically increasing the number of atoms in the active region.\cite{sanchez2010cholesky} Therefore, the results reported here are chosen only to demonstrate the improved representation given by MLHF-A/AB as compared to MLHF (Cholesky).

\subsection{Summary and Conclusions}

To summarize, we have presented a novel energy-based criterion to localize MOs in specific spatial regions of a molecular system. In particular, this approach is based on a MLHF partitioning of the system. Differently from other fragmentation methods, it is entirely based on HF theory, thus no approximations are introduced in the interaction energy. The prospects of our approach are demonstrated for four selected systems, characterized by both conjugated and non-conjugated skeletons. In particular, we have shown that MLHF-AB approach provides continuous PES, thus solving the discontinuity issues that can arise by exploiting common localization procedures in projection-based approaches.

The accuracy of our approach is then shown for ground state properties (dipole moments) and excitation energies calculated at the full CC2 level. The computational cost is reduced due to the partitioning of the system in active and inactive fragments. Both MLHF-A and MLHF-AB are able to reduce the discrepancy between MLHF and reference full coupled cluster values, in this way demonstrating their reliability in describing local excitations. Notice that in the present study the procedure is applied to relatively small molecules in order to allow a direct comparison with full coupled cluster results. However, the model has the potential to be applied to very large systems. A detailed benchmark of the performances of MLHF-A/AB on excitation energies will be the topic of future communications.

To conclude, the MO localization provided by our approach can have different applications, ranging from those illustrated in this work (i.e. local ground state properties\cite{giovannini2017disrep,giovannini2019eprdisrep} and local excitations\cite{hoyvik2017correlated,azarias2017modeling,giovannini2019quantum}) to the accurate calculation of interaction and reaction energies of molecular systems in large biological matrices or adsorbed on nanomaterials.\cite{khaliullin2007unravelling,su2009energy,boulanger2014toward} In addition, the local MOs obtained through our procedure may be used to define the different fragment densities in fragmentation approaches,\cite{gordon2012fragmentation} and different boundaries in the cap regions for QM/MM approaches when covalent bonds are cut.\cite{senn2009qm} 

\section*{Supporting Information}

Data related to Figs. 2-5 and 7.

\section*{Acknowledgments}

We acknowledge Sarai Dery Folkestad and Ida-Marie H{\o}yvik (NTNU) for discussions on technical aspects of the implementation. We acknowledge Chiara Cappelli (SNS) for computer resources. 
We acknowledge funding from the Marie Sklodowska-Curie European Training Network “COSINE - COmputational Spectroscopy In Natural sciences and Engineering”, Grant Agreement No.  765739, and the Research Council of Norway through FRINATEK projects 263110 and 275506.


{
\small
\bibliography{biblio}

\providecommand{\latin}[1]{#1}
\makeatletter
\providecommand{\doi}
  {\begingroup\let\do\@makeother\dospecials
  \catcode`\{=1 \catcode`\}=2 \doi@aux}
\providecommand{\doi@aux}[1]{\endgroup\texttt{#1}}
\makeatother
\providecommand*\mcitethebibliography{\thebibliography}
\csname @ifundefined\endcsname{endmcitethebibliography}
  {\let\endmcitethebibliography\endthebibliography}{}
\begin{mcitethebibliography}{73}
\providecommand*\natexlab[1]{#1}
\providecommand*\mciteSetBstSublistMode[1]{}
\providecommand*\mciteSetBstMaxWidthForm[2]{}
\providecommand*\mciteBstWouldAddEndPuncttrue
  {\def\EndOfBibitem{\unskip.}}
\providecommand*\mciteBstWouldAddEndPunctfalse
  {\let\EndOfBibitem\relax}
\providecommand*\mciteSetBstMidEndSepPunct[3]{}
\providecommand*\mciteSetBstSublistLabelBeginEnd[3]{}
\providecommand*\EndOfBibitem{}
\mciteSetBstSublistMode{f}
\mciteSetBstMaxWidthForm{subitem}{(\alph{mcitesubitemcount})}
\mciteSetBstSublistLabelBeginEnd
  {\mcitemaxwidthsubitemform\space}
  {\relax}
  {\relax}

\bibitem[H{\o}yvik and J{\o}rgensen(2016)H{\o}yvik, and
  J{\o}rgensen]{hoyvik2016characterization}
H{\o}yvik,~I.-M.; J{\o}rgensen,~P. Characterization and generation of local
  occupied and virtual Hartree--Fock orbitals. \emph{Chem. Rev.} \textbf{2016},
  \emph{116}, 3306--3327\relax
\mciteBstWouldAddEndPuncttrue
\mciteSetBstMidEndSepPunct{\mcitedefaultmidpunct}
{\mcitedefaultendpunct}{\mcitedefaultseppunct}\relax
\EndOfBibitem
\bibitem[Ma and Werner(2018)Ma, and Werner]{ma2018explicitly}
Ma,~Q.; Werner,~H.-J. Explicitly correlated local coupled-cluster methods using
  pair natural orbitals. \emph{WIREs Comput. Mol. Sci.} \textbf{2018},
  \emph{8}, e1371\relax
\mciteBstWouldAddEndPuncttrue
\mciteSetBstMidEndSepPunct{\mcitedefaultmidpunct}
{\mcitedefaultendpunct}{\mcitedefaultseppunct}\relax
\EndOfBibitem
\bibitem[Edmiston and Ruedenberg(1963)Edmiston, and
  Ruedenberg]{edmiston1963localized}
Edmiston,~C.; Ruedenberg,~K. Localized atomic and molecular orbitals.
  \emph{Rev. Mod. Phys.} \textbf{1963}, \emph{35}, 457\relax
\mciteBstWouldAddEndPuncttrue
\mciteSetBstMidEndSepPunct{\mcitedefaultmidpunct}
{\mcitedefaultendpunct}{\mcitedefaultseppunct}\relax
\EndOfBibitem
\bibitem[Boughton and Pulay(1993)Boughton, and Pulay]{boughton1993comparison}
Boughton,~J.~W.; Pulay,~P. Comparison of the boys and Pipek--Mezey
  localizations in the local correlation approach and automatic virtual basis
  selection. \emph{J. Comput. Chem.} \textbf{1993}, \emph{14}, 736--740\relax
\mciteBstWouldAddEndPuncttrue
\mciteSetBstMidEndSepPunct{\mcitedefaultmidpunct}
{\mcitedefaultendpunct}{\mcitedefaultseppunct}\relax
\EndOfBibitem
\bibitem[Khaliullin \latin{et~al.}(2008)Khaliullin, Bell, and
  Head-Gordon]{khaliullin2008analysis}
Khaliullin,~R.~Z.; Bell,~A.~T.; Head-Gordon,~M. Analysis of charge transfer
  effects in molecular complexes based on absolutely localized molecular
  orbitals. \emph{J. Chem. Phys.} \textbf{2008}, \emph{128}, 184112\relax
\mciteBstWouldAddEndPuncttrue
\mciteSetBstMidEndSepPunct{\mcitedefaultmidpunct}
{\mcitedefaultendpunct}{\mcitedefaultseppunct}\relax
\EndOfBibitem
\bibitem[Khaliullin \latin{et~al.}(2007)Khaliullin, Cobar, Lochan, Bell, and
  Head-Gordon]{khaliullin2007unravelling}
Khaliullin,~R.~Z.; Cobar,~E.~A.; Lochan,~R.~C.; Bell,~A.~T.; Head-Gordon,~M.
  Unravelling the origin of intermolecular interactions using absolutely
  localized molecular orbitals. \emph{J. Phys. Chem. A} \textbf{2007},
  \emph{111}, 8753--8765\relax
\mciteBstWouldAddEndPuncttrue
\mciteSetBstMidEndSepPunct{\mcitedefaultmidpunct}
{\mcitedefaultendpunct}{\mcitedefaultseppunct}\relax
\EndOfBibitem
\bibitem[Aquilante \latin{et~al.}(2006)Aquilante, Bondo~Pedersen,
  S{\'a}nchez~de Mer{\'a}s, and Koch]{aquilante2006fast}
Aquilante,~F.; Bondo~Pedersen,~T.; S{\'a}nchez~de Mer{\'a}s,~A.; Koch,~H. Fast
  noniterative orbital localization for large molecules. \emph{J. Chem. Phys.}
  \textbf{2006}, \emph{125}, 174101\relax
\mciteBstWouldAddEndPuncttrue
\mciteSetBstMidEndSepPunct{\mcitedefaultmidpunct}
{\mcitedefaultendpunct}{\mcitedefaultseppunct}\relax
\EndOfBibitem
\bibitem[H{\o}yvik \latin{et~al.}(2012)H{\o}yvik, Jansik, and
  J{\o}rgensen]{hoyvik2012orbital}
H{\o}yvik,~I.-M.; Jansik,~B.; J{\o}rgensen,~P. Orbital localization using
  fourth central moment minimization. \emph{J. Chem. Phys.} \textbf{2012},
  \emph{137}, 224114\relax
\mciteBstWouldAddEndPuncttrue
\mciteSetBstMidEndSepPunct{\mcitedefaultmidpunct}
{\mcitedefaultendpunct}{\mcitedefaultseppunct}\relax
\EndOfBibitem
\bibitem[Jans{\'\i}k \latin{et~al.}(2011)Jans{\'\i}k, H{\o}st, Kristensen, and
  J{\o}rgensen]{jansik2011local}
Jans{\'\i}k,~B.; H{\o}st,~S.; Kristensen,~K.; J{\o}rgensen,~P. Local orbitals
  by minimizing powers of the orbital variance. \emph{J. Chem. Phys.}
  \textbf{2011}, \emph{134}, 194104\relax
\mciteBstWouldAddEndPuncttrue
\mciteSetBstMidEndSepPunct{\mcitedefaultmidpunct}
{\mcitedefaultendpunct}{\mcitedefaultseppunct}\relax
\EndOfBibitem
\bibitem[H{\o}yvik \latin{et~al.}(2015)H{\o}yvik, Kristensen, Kj{\ae}rgaard,
  and J{\o}rgensen]{hoyvik2015perspective}
H{\o}yvik,~I.-M.; Kristensen,~K.; Kj{\ae}rgaard,~T.; J{\o}rgensen,~P.
  \emph{Thom H. Dunning, Jr.}; Springer, 2015; pp 287--296\relax
\mciteBstWouldAddEndPuncttrue
\mciteSetBstMidEndSepPunct{\mcitedefaultmidpunct}
{\mcitedefaultendpunct}{\mcitedefaultseppunct}\relax
\EndOfBibitem
\bibitem[H{\o}yvik \latin{et~al.}(2012)H{\o}yvik, Jansik, and
  J{\o}rgensen]{hoyvik2012trust}
H{\o}yvik,~I.-M.; Jansik,~B.; J{\o}rgensen,~P. Trust region minimization of
  orbital localization functions. \emph{J. Chem. Theory Comput.} \textbf{2012},
  \emph{8}, 3137--3146\relax
\mciteBstWouldAddEndPuncttrue
\mciteSetBstMidEndSepPunct{\mcitedefaultmidpunct}
{\mcitedefaultendpunct}{\mcitedefaultseppunct}\relax
\EndOfBibitem
\bibitem[Zi{\'o}{\l}kowski \latin{et~al.}(2009)Zi{\'o}{\l}kowski, Jansik,
  J{\o}rgensen, and Olsen]{ziolkowski2009maximum}
Zi{\'o}{\l}kowski,~M.; Jansik,~B.; J{\o}rgensen,~P.; Olsen,~J. Maximum locality
  in occupied and virtual orbital spaces using a least-change strategy.
  \emph{J. Chem. Phys.} \textbf{2009}, \emph{131}, 124112\relax
\mciteBstWouldAddEndPuncttrue
\mciteSetBstMidEndSepPunct{\mcitedefaultmidpunct}
{\mcitedefaultendpunct}{\mcitedefaultseppunct}\relax
\EndOfBibitem
\bibitem[Gianinetti \latin{et~al.}(1996)Gianinetti, Raimondi, and
  Tornaghi]{gianinetti1996modification}
Gianinetti,~E.; Raimondi,~M.; Tornaghi,~E. Modification of the Roothaan
  equations to exclude BSSE from molecular interaction calculations. \emph{Int.
  J. Quantum Chem.} \textbf{1996}, \emph{60}, 157--166\relax
\mciteBstWouldAddEndPuncttrue
\mciteSetBstMidEndSepPunct{\mcitedefaultmidpunct}
{\mcitedefaultendpunct}{\mcitedefaultseppunct}\relax
\EndOfBibitem
\bibitem[Stoll \latin{et~al.}(1980)Stoll, Wagenblast, and
  Preu$\beta$]{stoll1980use}
Stoll,~H.; Wagenblast,~G.; Preu$\beta$,~H. On the use of local basis sets for
  localized molecular orbitals. \emph{Theor. Chim. Acta} \textbf{1980},
  \emph{57}, 169--178\relax
\mciteBstWouldAddEndPuncttrue
\mciteSetBstMidEndSepPunct{\mcitedefaultmidpunct}
{\mcitedefaultendpunct}{\mcitedefaultseppunct}\relax
\EndOfBibitem
\bibitem[Li \latin{et~al.}(2016)Li, Ni, and Li]{li2016cluster}
Li,~W.; Ni,~Z.; Li,~S. Cluster-in-molecule local correlation method for
  post-Hartree--Fock calculations of large systems. \emph{Mol. Phys.}
  \textbf{2016}, \emph{114}, 1447--1460\relax
\mciteBstWouldAddEndPuncttrue
\mciteSetBstMidEndSepPunct{\mcitedefaultmidpunct}
{\mcitedefaultendpunct}{\mcitedefaultseppunct}\relax
\EndOfBibitem
\bibitem[Azarias \latin{et~al.}(2017)Azarias, Russo, Cupellini, Mennucci, and
  Jacquemin]{azarias2017modeling}
Azarias,~C.; Russo,~R.; Cupellini,~L.; Mennucci,~B.; Jacquemin,~D. Modeling
  excitation energy transfer in multi-BODIPY architectures. \emph{Phys. Chem.
  Chem. Phys.} \textbf{2017}, \emph{19}, 6443--6453\relax
\mciteBstWouldAddEndPuncttrue
\mciteSetBstMidEndSepPunct{\mcitedefaultmidpunct}
{\mcitedefaultendpunct}{\mcitedefaultseppunct}\relax
\EndOfBibitem
\bibitem[Mennucci and Corni(2019)Mennucci, and Corni]{mennucci2019multiscale}
Mennucci,~B.; Corni,~S. Multiscale modelling of photoinduced processes in
  composite systems. \emph{Nat. Rev. Chem.} \textbf{2019}, \emph{3},
  315–330\relax
\mciteBstWouldAddEndPuncttrue
\mciteSetBstMidEndSepPunct{\mcitedefaultmidpunct}
{\mcitedefaultendpunct}{\mcitedefaultseppunct}\relax
\EndOfBibitem
\bibitem[Warshel and Levitt(1976)Warshel, and Levitt]{warshel1976theoretical}
Warshel,~A.; Levitt,~M. Theoretical studies of enzymic reactions: dielectric,
  electrostatic and steric stabilization of the carbonium ion in the reaction
  of lysozyme. \emph{J. Mol. Biol.} \textbf{1976}, \emph{103}, 227--249\relax
\mciteBstWouldAddEndPuncttrue
\mciteSetBstMidEndSepPunct{\mcitedefaultmidpunct}
{\mcitedefaultendpunct}{\mcitedefaultseppunct}\relax
\EndOfBibitem
\bibitem[Cappelli(2016)]{cappelli2016integrated}
Cappelli,~C. Integrated QM/Polarizable MM/Continuum Approaches to Model
  Chiroptical Properties of Strongly Interacting Solute-Solvent Systems.
  \emph{Int. J. Quantum Chem.} \textbf{2016}, \emph{116}, 1532--1542\relax
\mciteBstWouldAddEndPuncttrue
\mciteSetBstMidEndSepPunct{\mcitedefaultmidpunct}
{\mcitedefaultendpunct}{\mcitedefaultseppunct}\relax
\EndOfBibitem
\bibitem[Mennucci(2012)]{Mennucci12_386}
Mennucci,~B. Polarizable Continuum Model. \emph{WIREs Comput. Mol. Sci.}
  \textbf{2012}, \emph{2}, 386--404\relax
\mciteBstWouldAddEndPuncttrue
\mciteSetBstMidEndSepPunct{\mcitedefaultmidpunct}
{\mcitedefaultendpunct}{\mcitedefaultseppunct}\relax
\EndOfBibitem
\bibitem[Giovannini \latin{et~al.}(2019)Giovannini, Puglisi, Ambrosetti, and
  Cappelli]{giovannini2019fqfmu}
Giovannini,~T.; Puglisi,~A.; Ambrosetti,~M.; Cappelli,~C. Polarizable QM/MM
  approach with fluctuating charges and fluctuating dipoles: the QM/FQF$\mu$
  model. \emph{J. Chem. Theory Comput.} \textbf{2019}, \emph{15},
  2233--2245\relax
\mciteBstWouldAddEndPuncttrue
\mciteSetBstMidEndSepPunct{\mcitedefaultmidpunct}
{\mcitedefaultendpunct}{\mcitedefaultseppunct}\relax
\EndOfBibitem
\bibitem[Gordon \latin{et~al.}(2012)Gordon, Fedorov, Pruitt, and
  Slipchenko]{gordon2012fragmentation}
Gordon,~M.~S.; Fedorov,~D.~G.; Pruitt,~S.~R.; Slipchenko,~L.~V. Fragmentation
  methods: A route to accurate calculations on large systems. \emph{Chem. Rev.}
  \textbf{2012}, \emph{112}, 632--672\relax
\mciteBstWouldAddEndPuncttrue
\mciteSetBstMidEndSepPunct{\mcitedefaultmidpunct}
{\mcitedefaultendpunct}{\mcitedefaultseppunct}\relax
\EndOfBibitem
\bibitem[Collins and Bettens(2015)Collins, and Bettens]{collins2015energy}
Collins,~M.~A.; Bettens,~R.~P. Energy-based molecular fragmentation methods.
  \emph{Chem. Rev.} \textbf{2015}, \emph{115}, 5607--5642\relax
\mciteBstWouldAddEndPuncttrue
\mciteSetBstMidEndSepPunct{\mcitedefaultmidpunct}
{\mcitedefaultendpunct}{\mcitedefaultseppunct}\relax
\EndOfBibitem
\bibitem[Pruitt \latin{et~al.}(2014)Pruitt, Bertoni, Brorsen, and
  Gordon]{pruitt2014efficient}
Pruitt,~S.~R.; Bertoni,~C.; Brorsen,~K.~R.; Gordon,~M.~S. Efficient and
  accurate fragmentation methods. \emph{Acc. Chem. Res.} \textbf{2014},
  \emph{47}, 2786--2794\relax
\mciteBstWouldAddEndPuncttrue
\mciteSetBstMidEndSepPunct{\mcitedefaultmidpunct}
{\mcitedefaultendpunct}{\mcitedefaultseppunct}\relax
\EndOfBibitem
\bibitem[Collins \latin{et~al.}(2014)Collins, Cvitkovic, and
  Bettens]{collins2014combined}
Collins,~M.~A.; Cvitkovic,~M.~W.; Bettens,~R.~P. The combined fragmentation and
  systematic molecular fragmentation methods. \emph{Acc. Chem. Res.}
  \textbf{2014}, \emph{47}, 2776--2785\relax
\mciteBstWouldAddEndPuncttrue
\mciteSetBstMidEndSepPunct{\mcitedefaultmidpunct}
{\mcitedefaultendpunct}{\mcitedefaultseppunct}\relax
\EndOfBibitem
\bibitem[Pruitt \latin{et~al.}(2012)Pruitt, Addicoat, Collins, and
  Gordon]{pruitt2012fragment}
Pruitt,~S.~R.; Addicoat,~M.~A.; Collins,~M.~A.; Gordon,~M.~S. The fragment
  molecular orbital and systematic molecular fragmentation methods applied to
  water clusters. \emph{Phys. Chem. Chem. Phys.} \textbf{2012}, \emph{14},
  7752--7764\relax
\mciteBstWouldAddEndPuncttrue
\mciteSetBstMidEndSepPunct{\mcitedefaultmidpunct}
{\mcitedefaultendpunct}{\mcitedefaultseppunct}\relax
\EndOfBibitem
\bibitem[Ding \latin{et~al.}(2017)Ding, Manby, and
  Miller~III]{ding2017embedded}
Ding,~F.; Manby,~F.~R.; Miller~III,~T.~F. Embedded mean-field theory with
  block-orthogonalized partitioning. \emph{J. Chem. Theory Comput.}
  \textbf{2017}, \emph{13}, 1605--1615\relax
\mciteBstWouldAddEndPuncttrue
\mciteSetBstMidEndSepPunct{\mcitedefaultmidpunct}
{\mcitedefaultendpunct}{\mcitedefaultseppunct}\relax
\EndOfBibitem
\bibitem[Wen \latin{et~al.}(2020)Wen, Graham, Chulhai, and
  Goodpaster]{wen2019absolutely}
Wen,~X.; Graham,~D.~S.; Chulhai,~D.~V.; Goodpaster,~J.~D. Absolutely Localized
  Projection-Based Embedding for Excited States. \emph{J. Chem. Theory Comput.}
  \textbf{2020}, \emph{16}, 385--398\relax
\mciteBstWouldAddEndPuncttrue
\mciteSetBstMidEndSepPunct{\mcitedefaultmidpunct}
{\mcitedefaultendpunct}{\mcitedefaultseppunct}\relax
\EndOfBibitem
\bibitem[Bennie \latin{et~al.}(2017)Bennie, Curchod, Manby, and
  Glowacki]{bennie2017pushing}
Bennie,~S.~J.; Curchod,~B.~F.; Manby,~F.~R.; Glowacki,~D.~R. Pushing the limits
  of EOM-CCSD with projector-based embedding for excitation energies. \emph{J.
  Phys. Chem. Lett.} \textbf{2017}, \emph{8}, 5559--5565\relax
\mciteBstWouldAddEndPuncttrue
\mciteSetBstMidEndSepPunct{\mcitedefaultmidpunct}
{\mcitedefaultendpunct}{\mcitedefaultseppunct}\relax
\EndOfBibitem
\bibitem[Chen and Gao(2020)Chen, and Gao]{chen2020fragment}
Chen,~X.; Gao,~J. Fragment Exchange Potential for Realizing Pauli Deformation
  of Inter-Fragment Interactions. \emph{J. Phys. Chem. Lett.} \textbf{2020},
  10.1021/acs.jpclett.0c00933\relax
\mciteBstWouldAddEndPuncttrue
\mciteSetBstMidEndSepPunct{\mcitedefaultmidpunct}
{\mcitedefaultendpunct}{\mcitedefaultseppunct}\relax
\EndOfBibitem
\bibitem[Boys(1960)]{boys1960construction}
Boys,~S.~F. Construction of some molecular orbitals to be approximately
  invariant for changes from one molecule to another. \emph{Rev. Mod. Phys.}
  \textbf{1960}, \emph{32}, 296\relax
\mciteBstWouldAddEndPuncttrue
\mciteSetBstMidEndSepPunct{\mcitedefaultmidpunct}
{\mcitedefaultendpunct}{\mcitedefaultseppunct}\relax
\EndOfBibitem
\bibitem[Chulhai and Goodpaster(2017)Chulhai, and
  Goodpaster]{chulhai2017improved}
Chulhai,~D.~V.; Goodpaster,~J.~D. Improved accuracy and efficiency in quantum
  embedding through absolute localization. \emph{J. Chem. Theory Comput.}
  \textbf{2017}, \emph{13}, 1503--1508\relax
\mciteBstWouldAddEndPuncttrue
\mciteSetBstMidEndSepPunct{\mcitedefaultmidpunct}
{\mcitedefaultendpunct}{\mcitedefaultseppunct}\relax
\EndOfBibitem
\bibitem[Chulhai and Goodpaster(2018)Chulhai, and
  Goodpaster]{chulhai2018projection}
Chulhai,~D.~V.; Goodpaster,~J.~D. Projection-based correlated wave function in
  density functional theory embedding for periodic systems. \emph{J. Chem.
  Theory Comput.} \textbf{2018}, \emph{14}, 1928--1942\relax
\mciteBstWouldAddEndPuncttrue
\mciteSetBstMidEndSepPunct{\mcitedefaultmidpunct}
{\mcitedefaultendpunct}{\mcitedefaultseppunct}\relax
\EndOfBibitem
\bibitem[Sayfutyarova \latin{et~al.}(2017)Sayfutyarova, Sun, Chan, and
  Knizia]{sayfutyarova2017automated}
Sayfutyarova,~E.~R.; Sun,~Q.; Chan,~G. K.-L.; Knizia,~G. Automated construction
  of molecular active spaces from atomic valence orbitals. \emph{J. Chem.
  Theory Comput.} \textbf{2017}, \emph{13}, 4063--4078\relax
\mciteBstWouldAddEndPuncttrue
\mciteSetBstMidEndSepPunct{\mcitedefaultmidpunct}
{\mcitedefaultendpunct}{\mcitedefaultseppunct}\relax
\EndOfBibitem
\bibitem[S{\ae}ther \latin{et~al.}(2017)S{\ae}ther, Kj{\ae}rgaard, Koch, and
  H{\o}yvik]{saether2017density}
S{\ae}ther,~S.; Kj{\ae}rgaard,~T.; Koch,~H.; H{\o}yvik,~I.-M. Density-Based
  Multilevel Hartree--Fock Model. \emph{J. Chem. Theory Comput.} \textbf{2017},
  \emph{13}, 5282--5290\relax
\mciteBstWouldAddEndPuncttrue
\mciteSetBstMidEndSepPunct{\mcitedefaultmidpunct}
{\mcitedefaultendpunct}{\mcitedefaultseppunct}\relax
\EndOfBibitem
\bibitem[H{\o}yvik(2020)]{hoyvik2020convergence}
H{\o}yvik,~I.-M. Convergence acceleration for the multilevel Hartree--Fock
  model. \emph{Mol. Phys.} \textbf{2020}, \emph{118}, 1626929\relax
\mciteBstWouldAddEndPuncttrue
\mciteSetBstMidEndSepPunct{\mcitedefaultmidpunct}
{\mcitedefaultendpunct}{\mcitedefaultseppunct}\relax
\EndOfBibitem
\bibitem[Culpitt \latin{et~al.}(2017)Culpitt, Brorsen, and
  Hammes-Schiffer]{culpitt2017communication}
Culpitt,~T.; Brorsen,~K.~R.; Hammes-Schiffer,~S. Communication: Density
  functional theory embedding with the orthogonality constrained basis set
  expansion procedure. \emph{J. Chem. Phys.} \textbf{2017}, \emph{146},
  211101\relax
\mciteBstWouldAddEndPuncttrue
\mciteSetBstMidEndSepPunct{\mcitedefaultmidpunct}
{\mcitedefaultendpunct}{\mcitedefaultseppunct}\relax
\EndOfBibitem
\bibitem[H{\'e}gely \latin{et~al.}(2016)H{\'e}gely, Nagy, Ferenczy, and
  K{\'a}llay]{hegely2016exact}
H{\'e}gely,~B.; Nagy,~P.~R.; Ferenczy,~G.~G.; K{\'a}llay,~M. Exact density
  functional and wave function embedding schemes based on orbital localization.
  \emph{J. Chem. Phys.} \textbf{2016}, \emph{145}, 064107\relax
\mciteBstWouldAddEndPuncttrue
\mciteSetBstMidEndSepPunct{\mcitedefaultmidpunct}
{\mcitedefaultendpunct}{\mcitedefaultseppunct}\relax
\EndOfBibitem
\bibitem[Huzinaga and Cantu(1971)Huzinaga, and Cantu]{huzinaga1971theory}
Huzinaga,~S.; Cantu,~A. Theory of separability of many-electron systems.
  \emph{J. Chem. Phys.} \textbf{1971}, \emph{55}, 5543--5549\relax
\mciteBstWouldAddEndPuncttrue
\mciteSetBstMidEndSepPunct{\mcitedefaultmidpunct}
{\mcitedefaultendpunct}{\mcitedefaultseppunct}\relax
\EndOfBibitem
\bibitem[Goodpaster \latin{et~al.}(2010)Goodpaster, Ananth, Manby, and
  Miller~III]{goodpaster2010exact}
Goodpaster,~J.~D.; Ananth,~N.; Manby,~F.~R.; Miller~III,~T.~F. Exact
  nonadditive kinetic potentials for embedded density functional theory.
  \emph{J. Chem. Phys.} \textbf{2010}, \emph{133}, 084103\relax
\mciteBstWouldAddEndPuncttrue
\mciteSetBstMidEndSepPunct{\mcitedefaultmidpunct}
{\mcitedefaultendpunct}{\mcitedefaultseppunct}\relax
\EndOfBibitem
\bibitem[Goodpaster \latin{et~al.}(2012)Goodpaster, Barnes, Manby, and
  Miller~III]{goodpaster2012density}
Goodpaster,~J.~D.; Barnes,~T.~A.; Manby,~F.~R.; Miller~III,~T.~F. Density
  functional theory embedding for correlated wavefunctions: Improved methods
  for open-shell systems and transition metal complexes. \emph{J. Chem. Phys.}
  \textbf{2012}, \emph{137}, 224113\relax
\mciteBstWouldAddEndPuncttrue
\mciteSetBstMidEndSepPunct{\mcitedefaultmidpunct}
{\mcitedefaultendpunct}{\mcitedefaultseppunct}\relax
\EndOfBibitem
\bibitem[Goodpaster \latin{et~al.}(2014)Goodpaster, Barnes, Manby, and
  Miller~III]{goodpaster2014accurate}
Goodpaster,~J.~D.; Barnes,~T.~A.; Manby,~F.~R.; Miller~III,~T.~F. Accurate and
  systematically improvable density functional theory embedding for correlated
  wavefunctions. \emph{J. Chem. Phys.} \textbf{2014}, \emph{140}, 18A507\relax
\mciteBstWouldAddEndPuncttrue
\mciteSetBstMidEndSepPunct{\mcitedefaultmidpunct}
{\mcitedefaultendpunct}{\mcitedefaultseppunct}\relax
\EndOfBibitem
\bibitem[Manby \latin{et~al.}(2012)Manby, Stella, Goodpaster, and
  Miller~III]{manby2012simple}
Manby,~F.~R.; Stella,~M.; Goodpaster,~J.~D.; Miller~III,~T.~F. A simple, exact
  density-functional-theory embedding scheme. \emph{J. Chem. Theory Comput.}
  \textbf{2012}, \emph{8}, 2564--2568\relax
\mciteBstWouldAddEndPuncttrue
\mciteSetBstMidEndSepPunct{\mcitedefaultmidpunct}
{\mcitedefaultendpunct}{\mcitedefaultseppunct}\relax
\EndOfBibitem
\bibitem[Sun and Chan(2016)Sun, and Chan]{sun2016quantum}
Sun,~Q.; Chan,~G. K.-L. Quantum embedding theories. \emph{Acc. Chem. Res.}
  \textbf{2016}, \emph{49}, 2705--2712\relax
\mciteBstWouldAddEndPuncttrue
\mciteSetBstMidEndSepPunct{\mcitedefaultmidpunct}
{\mcitedefaultendpunct}{\mcitedefaultseppunct}\relax
\EndOfBibitem
\bibitem[Myhre and Koch(2016)Myhre, and Koch]{myhre2016multilevel}
Myhre,~R.~H.; Koch,~H. The multilevel CC3 coupled cluster model. \emph{J. Chem.
  Phys.} \textbf{2016}, \emph{145}, 044111\relax
\mciteBstWouldAddEndPuncttrue
\mciteSetBstMidEndSepPunct{\mcitedefaultmidpunct}
{\mcitedefaultendpunct}{\mcitedefaultseppunct}\relax
\EndOfBibitem
\bibitem[Folkestad and Koch(2019)Folkestad, and Koch]{folkestad2019multilevel}
Folkestad,~S.~D.; Koch,~H. Multilevel CC2 and CCSD Methods with Correlated
  Natural Transition Orbitals. \emph{J. Chem. Theory Comput.} \textbf{2019},
  \relax
\mciteBstWouldAddEndPunctfalse
\mciteSetBstMidEndSepPunct{\mcitedefaultmidpunct}
{}{\mcitedefaultseppunct}\relax
\EndOfBibitem
\bibitem[Folkestad \latin{et~al.}(2020)Folkestad, Kjønstad, Myhre, Andersen,
  Balbi, Coriani, Giovannini, Goletto, Haugland, Hutcheson, Høyvik, Moitra,
  Paul, Scavino, Skeidsvoll, Åsmund H.~Tveten, and Koch]{eT_arxiv}
Folkestad,~S.~D.; Kjønstad,~E.~F.; Myhre,~R.~H.; Andersen,~J.~H.; Balbi,~A.;
  Coriani,~S.; Giovannini,~T.; Goletto,~L.; Haugland,~T.~S.; Hutcheson,~A.;
  Høyvik,~I.-M.; Moitra,~T.; Paul,~A.~C.; Scavino,~M.; Skeidsvoll,~A.~S.;
  Åsmund H.~Tveten,; Koch,~H. eT 1.0: an open source electronic structure
  program with emphasis on coupled cluster and multilevel methods. \emph{arXiv}
  \textbf{2020}, 2002.05631\relax
\mciteBstWouldAddEndPuncttrue
\mciteSetBstMidEndSepPunct{\mcitedefaultmidpunct}
{\mcitedefaultendpunct}{\mcitedefaultseppunct}\relax
\EndOfBibitem
\bibitem[Aquilante \latin{et~al.}(2011)Aquilante, Boman, Bostr{\"o}m, Koch,
  Lindh, de~Mer{\'a}s, and Pedersen]{aquilante2011cholesky}
Aquilante,~F.; Boman,~L.; Bostr{\"o}m,~J.; Koch,~H.; Lindh,~R.;
  de~Mer{\'a}s,~A.~S.; Pedersen,~T.~B. \emph{Linear-Scaling Techniques in
  Computational Chemistry and Physics}; Springer, 2011; pp 301--343\relax
\mciteBstWouldAddEndPuncttrue
\mciteSetBstMidEndSepPunct{\mcitedefaultmidpunct}
{\mcitedefaultendpunct}{\mcitedefaultseppunct}\relax
\EndOfBibitem
\bibitem[S{\'a}nchez~de Mer{\'a}s \latin{et~al.}(2010)S{\'a}nchez~de Mer{\'a}s,
  Koch, Cuesta, and Boman]{sanchez2010cholesky}
S{\'a}nchez~de Mer{\'a}s,~A.~M.; Koch,~H.; Cuesta,~I.~G.; Boman,~L. Cholesky
  decomposition-based definition of atomic subsystems in electronic structure
  calculations. \emph{J. Chem. Phys.} \textbf{2010}, \emph{132}, 204105\relax
\mciteBstWouldAddEndPuncttrue
\mciteSetBstMidEndSepPunct{\mcitedefaultmidpunct}
{\mcitedefaultendpunct}{\mcitedefaultseppunct}\relax
\EndOfBibitem
\bibitem[Koch \latin{et~al.}(2003)Koch, S{\'a}nchez~de Mer{\'a}s, and
  Pedersen]{koch2003reduced}
Koch,~H.; S{\'a}nchez~de Mer{\'a}s,~A.; Pedersen,~T.~B. Reduced scaling in
  electronic structure calculations using Cholesky decompositions. \emph{J.
  Chem. Phys.} \textbf{2003}, \emph{118}, 9481--9484\relax
\mciteBstWouldAddEndPuncttrue
\mciteSetBstMidEndSepPunct{\mcitedefaultmidpunct}
{\mcitedefaultendpunct}{\mcitedefaultseppunct}\relax
\EndOfBibitem
\bibitem[Christiansen \latin{et~al.}(2006)Christiansen, Manninen, J{\o}rgensen,
  and Olsen]{christiansen2006coupled}
Christiansen,~O.; Manninen,~P.; J{\o}rgensen,~P.; Olsen,~J. Coupled-cluster
  theory in a projected atomic orbital basis. \emph{J. Chem. Phys.}
  \textbf{2006}, \emph{124}, 084103\relax
\mciteBstWouldAddEndPuncttrue
\mciteSetBstMidEndSepPunct{\mcitedefaultmidpunct}
{\mcitedefaultendpunct}{\mcitedefaultseppunct}\relax
\EndOfBibitem
\bibitem[Myhre \latin{et~al.}(2014)Myhre, S{\'a}nchez~de Mer{\'a}s, and
  Koch]{myhre2014multi}
Myhre,~R.~H.; S{\'a}nchez~de Mer{\'a}s,~A.~M.; Koch,~H. Multi-level coupled
  cluster theory. \emph{J. Chem. Phys.} \textbf{2014}, \emph{141}, 224105\relax
\mciteBstWouldAddEndPuncttrue
\mciteSetBstMidEndSepPunct{\mcitedefaultmidpunct}
{\mcitedefaultendpunct}{\mcitedefaultseppunct}\relax
\EndOfBibitem
\bibitem[Egidi \latin{et~al.}(2014)Egidi, Segado, Koch, Cappelli, and
  Barone]{egidi2014benchmark}
Egidi,~F.; Segado,~M.; Koch,~H.; Cappelli,~C.; Barone,~V. A benchmark study of
  electronic excitation energies, transition moments, and excited-state energy
  gradients on the nicotine molecule. \emph{J. Chem. Phys.} \textbf{2014},
  \emph{141}, 224114\relax
\mciteBstWouldAddEndPuncttrue
\mciteSetBstMidEndSepPunct{\mcitedefaultmidpunct}
{\mcitedefaultendpunct}{\mcitedefaultseppunct}\relax
\EndOfBibitem
\bibitem[Marder \latin{et~al.}(1991)Marder, Beratan, and
  Cheng]{marder1991approaches}
Marder,~S.; Beratan,~D.; Cheng,~L.-T. Approaches for optimizing the first
  electronic hyperpolarizability of conjugated organic molecules.
  \emph{Science} \textbf{1991}, \emph{252}, 103--106\relax
\mciteBstWouldAddEndPuncttrue
\mciteSetBstMidEndSepPunct{\mcitedefaultmidpunct}
{\mcitedefaultendpunct}{\mcitedefaultseppunct}\relax
\EndOfBibitem
\bibitem[Grigorenko \latin{et~al.}(2012)Grigorenko, Polini, and
  Novoselov]{grigorenko2012graphene}
Grigorenko,~A.; Polini,~M.; Novoselov,~K. Graphene plasmonics. \emph{Nat.
  Photonics} \textbf{2012}, \emph{6}, 749\relax
\mciteBstWouldAddEndPuncttrue
\mciteSetBstMidEndSepPunct{\mcitedefaultmidpunct}
{\mcitedefaultendpunct}{\mcitedefaultseppunct}\relax
\EndOfBibitem
\bibitem[Gibson and Knight(2003)Gibson, and Knight]{gibson20032}
Gibson,~S.~E.; Knight,~J.~D. [2.2] Paracyclophane derivatives in asymmetric
  catalysis. \emph{Org. Biomol. Chem.} \textbf{2003}, \emph{1},
  1256--1269\relax
\mciteBstWouldAddEndPuncttrue
\mciteSetBstMidEndSepPunct{\mcitedefaultmidpunct}
{\mcitedefaultendpunct}{\mcitedefaultseppunct}\relax
\EndOfBibitem
\bibitem[Gleiter and Hopf(2006)Gleiter, and Hopf]{gleiter2006modern}
Gleiter,~R.; Hopf,~H. \emph{Modern cyclophane chemistry}; John Wiley \& Sons,
  2006\relax
\mciteBstWouldAddEndPuncttrue
\mciteSetBstMidEndSepPunct{\mcitedefaultmidpunct}
{\mcitedefaultendpunct}{\mcitedefaultseppunct}\relax
\EndOfBibitem
\bibitem[Grimme(2004)]{grimme2004importance}
Grimme,~S. On the Importance of Electron Correlation Effects for the
  $\pi$-$\pi$ Interactions in Cyclophanes. \emph{Chem. Eur.} \textbf{2004},
  \emph{10}, 3423--3429\relax
\mciteBstWouldAddEndPuncttrue
\mciteSetBstMidEndSepPunct{\mcitedefaultmidpunct}
{\mcitedefaultendpunct}{\mcitedefaultseppunct}\relax
\EndOfBibitem
\bibitem[Demissie \latin{et~al.}(2016)Demissie, Dodziuk, Waluk, Ruud, Pietrzak,
  Vetokhina, Szyma{\`n}ski, Ja{\`z}wi{\'n}ski, and Hopf]{demissie2016pcp}
Demissie,~T.~B.; Dodziuk,~H.; Waluk,~J.; Ruud,~K.; Pietrzak,~M.; Vetokhina,~V.;
  Szyma{\`n}ski,~S.; Ja{\`z}wi{\'n}ski,~J.; Hopf,~H. Structure, NMR and
  Electronic Spectra of [m.n]Paracyclophanes with Varying Bridges Lengths (m, n
  = 2–4). \emph{J. Phys. Chem. A} \textbf{2016}, \emph{120}, 724--736\relax
\mciteBstWouldAddEndPuncttrue
\mciteSetBstMidEndSepPunct{\mcitedefaultmidpunct}
{\mcitedefaultendpunct}{\mcitedefaultseppunct}\relax
\EndOfBibitem
\bibitem[Bachrach(2011)]{bachrach2011pcp}
Bachrach,~S.~M. DFT Study of [2.2]-, [3.3]-, and [4.4]Paracyclophanes: Strain
  Energy, Conformations, and Rotational Barriers. \emph{J. Phys. Chem. A}
  \textbf{2011}, \emph{115}, 2396--2401\relax
\mciteBstWouldAddEndPuncttrue
\mciteSetBstMidEndSepPunct{\mcitedefaultmidpunct}
{\mcitedefaultendpunct}{\mcitedefaultseppunct}\relax
\EndOfBibitem
\bibitem[Frisch \latin{et~al.}(2016)Frisch, Trucks, Schlegel, Scuseria, Robb,
  Cheeseman, Scalmani, Barone, Petersson, Nakatsuji, Li, Caricato, Marenich,
  Bloino, Janesko, Gomperts, Mennucci, Hratchian, Ortiz, Izmaylov, Sonnenberg,
  Williams-Young, Ding, Lipparini, Egidi, Goings, Peng, Petrone, Henderson,
  Ranasinghe, Zakrzewski, Gao, Rega, Zheng, Liang, Hada, Ehara, Toyota, Fukuda,
  Hasegawa, Ishida, Nakajima, Honda, Kitao, Nakai, Vreven, Throssell,
  Montgomery, Peralta, Ogliaro, Bearpark, Heyd, Brothers, Kudin, Staroverov,
  Keith, Kobayashi, Normand, Raghavachari, Rendell, Burant, Iyengar, Tomasi,
  Cossi, Millam, Klene, Adamo, Cammi, Ochterski, Martin, Morokuma, Farkas,
  Foresman, and Fox]{gaussian16}
Frisch,~M.~J.; Trucks,~G.~W.; Schlegel,~H.~B.; Scuseria,~G.~E.; Robb,~M.~A.;
  Cheeseman,~J.~R.; Scalmani,~G.; Barone,~V.; Petersson,~G.~A.; Nakatsuji,~H.;
  Li,~X.; Caricato,~M.; Marenich,~A.~V.; Bloino,~J.; Janesko,~B.~G.;
  Gomperts,~R.; Mennucci,~B.; Hratchian,~H.~P.; Ortiz,~J.~V.; Izmaylov,~A.~F.;
  Sonnenberg,~J.~L.; Williams-Young,~D.; Ding,~F.; Lipparini,~F.; Egidi,~F.;
  Goings,~J.; Peng,~B.; Petrone,~A.; Henderson,~T.; Ranasinghe,~D.;
  Zakrzewski,~V.~G.; Gao,~J.; Rega,~N.; Zheng,~G.; Liang,~W.; Hada,~M.;
  Ehara,~M.; Toyota,~K.; Fukuda,~R.; Hasegawa,~J.; Ishida,~M.; Nakajima,~T.;
  Honda,~Y.; Kitao,~O.; Nakai,~H.; Vreven,~T.; Throssell,~K.;
  Montgomery,~J.~A.,~{Jr.}; Peralta,~J.~E.; Ogliaro,~F.; Bearpark,~M.~J.;
  Heyd,~J.~J.; Brothers,~E.~N.; Kudin,~K.~N.; Staroverov,~V.~N.; Keith,~T.~A.;
  Kobayashi,~R.; Normand,~J.; Raghavachari,~K.; Rendell,~A.~P.; Burant,~J.~C.;
  Iyengar,~S.~S.; Tomasi,~J.; Cossi,~M.; Millam,~J.~M.; Klene,~M.; Adamo,~C.;
  Cammi,~R.; Ochterski,~J.~W.; Martin,~R.~L.; Morokuma,~K.; Farkas,~O.;
  Foresman,~J.~B.; Fox,~D.~J. Gaussian~16 {R}evision {A}.03. 2016; Gaussian
  Inc. Wallingford CT\relax
\mciteBstWouldAddEndPuncttrue
\mciteSetBstMidEndSepPunct{\mcitedefaultmidpunct}
{\mcitedefaultendpunct}{\mcitedefaultseppunct}\relax
\EndOfBibitem
\bibitem[Castro-Neto \latin{et~al.}(2009)Castro-Neto, Guinea, Peres, Novoselov,
  and Geim]{neto2009electronic}
Castro-Neto,~A.~H.; Guinea,~F.; Peres,~N.~M.; Novoselov,~K.~S.; Geim,~A.~K. The
  electronic properties of graphene. \emph{Rev. Mod. Phys.} \textbf{2009},
  \emph{81}, 109\relax
\mciteBstWouldAddEndPuncttrue
\mciteSetBstMidEndSepPunct{\mcitedefaultmidpunct}
{\mcitedefaultendpunct}{\mcitedefaultseppunct}\relax
\EndOfBibitem
\bibitem[Zhang and Zhang(2003)Zhang, and Zhang]{zhang2003molecular}
Zhang,~D.~W.; Zhang,~J. Molecular fractionation with conjugate caps for full
  quantum mechanical calculation of protein--molecule interaction energy.
  \emph{J. Chem. Phys.} \textbf{2003}, \emph{119}, 3599--3605\relax
\mciteBstWouldAddEndPuncttrue
\mciteSetBstMidEndSepPunct{\mcitedefaultmidpunct}
{\mcitedefaultendpunct}{\mcitedefaultseppunct}\relax
\EndOfBibitem
\bibitem[Russ and Crawford(2004)Russ, and Crawford]{russ2004potential}
Russ,~N.~J.; Crawford,~T.~D. Potential energy surface discontinuities in local
  correlation methods. \emph{J. Chem. Phys.} \textbf{2004}, \emph{121},
  691--696\relax
\mciteBstWouldAddEndPuncttrue
\mciteSetBstMidEndSepPunct{\mcitedefaultmidpunct}
{\mcitedefaultendpunct}{\mcitedefaultseppunct}\relax
\EndOfBibitem
\bibitem[Christiansen \latin{et~al.}(1995)Christiansen, Koch, and
  J{\o}rgensen]{christiansen1995second}
Christiansen,~O.; Koch,~H.; J{\o}rgensen,~P. The second-order approximate
  coupled cluster singles and doubles model CC2. \emph{Chem. Phys. Lett.}
  \textbf{1995}, \emph{243}, 409--418\relax
\mciteBstWouldAddEndPuncttrue
\mciteSetBstMidEndSepPunct{\mcitedefaultmidpunct}
{\mcitedefaultendpunct}{\mcitedefaultseppunct}\relax
\EndOfBibitem
\bibitem[Giovannini \latin{et~al.}(2017)Giovannini, Lafiosca, and
  Cappelli]{giovannini2017disrep}
Giovannini,~T.; Lafiosca,~P.; Cappelli,~C. A General Route to Include Pauli
  Repulsion and Quantum Dispersion Effects in QM/MM Approaches. \emph{J. Chem.
  Theory Comput.} \textbf{2017}, \emph{13}, 4854--4870\relax
\mciteBstWouldAddEndPuncttrue
\mciteSetBstMidEndSepPunct{\mcitedefaultmidpunct}
{\mcitedefaultendpunct}{\mcitedefaultseppunct}\relax
\EndOfBibitem
\bibitem[Giovannini \latin{et~al.}(2019)Giovannini, Lafiosca, Chandramouli,
  Barone, and Cappelli]{giovannini2019eprdisrep}
Giovannini,~T.; Lafiosca,~P.; Chandramouli,~B.; Barone,~V.; Cappelli,~C.
  Effective yet Reliable Computation of Hyperfine Coupling Constants in
  Solution by a QM/MM Approach: Interplay Between Electrostatics and
  Non-electrostatic Effects. \emph{J. Chem. Phys.} \textbf{2019}, \emph{150},
  124102\relax
\mciteBstWouldAddEndPuncttrue
\mciteSetBstMidEndSepPunct{\mcitedefaultmidpunct}
{\mcitedefaultendpunct}{\mcitedefaultseppunct}\relax
\EndOfBibitem
\bibitem[Høyvik \latin{et~al.}(2017)Høyvik, Myhre, and
  Koch]{hoyvik2017correlated}
Høyvik,~I.-M.; Myhre,~R.~H.; Koch,~H. Correlated natural transition orbitals
  for core excitation energies in multilevel coupled cluster models. \emph{J.
  Chem. Phys.} \textbf{2017}, \emph{146}, 144109\relax
\mciteBstWouldAddEndPuncttrue
\mciteSetBstMidEndSepPunct{\mcitedefaultmidpunct}
{\mcitedefaultendpunct}{\mcitedefaultseppunct}\relax
\EndOfBibitem
\bibitem[Giovannini \latin{et~al.}(2019)Giovannini, Ambrosetti, and
  Cappelli]{giovannini2019quantum}
Giovannini,~T.; Ambrosetti,~M.; Cappelli,~C. Quantum Confinement Effects on
  Solvatochromic Shifts of Molecular Solutes. \emph{J. Phys. Chem. Lett.}
  \textbf{2019}, \emph{10}, 5823--5829\relax
\mciteBstWouldAddEndPuncttrue
\mciteSetBstMidEndSepPunct{\mcitedefaultmidpunct}
{\mcitedefaultendpunct}{\mcitedefaultseppunct}\relax
\EndOfBibitem
\bibitem[Su and Li(2009)Su, and Li]{su2009energy}
Su,~P.; Li,~H. Energy decomposition analysis of covalent bonds and
  intermolecular interactions. \emph{J. Chem.Phys.} \textbf{2009}, \emph{131},
  014102\relax
\mciteBstWouldAddEndPuncttrue
\mciteSetBstMidEndSepPunct{\mcitedefaultmidpunct}
{\mcitedefaultendpunct}{\mcitedefaultseppunct}\relax
\EndOfBibitem
\bibitem[Boulanger and Thiel(2014)Boulanger, and Thiel]{boulanger2014toward}
Boulanger,~E.; Thiel,~W. Toward QM/MM simulation of enzymatic reactions with
  the drude oscillator polarizable force field. \emph{J. Chem. Theory Comput.}
  \textbf{2014}, \emph{10}, 1795--1809\relax
\mciteBstWouldAddEndPuncttrue
\mciteSetBstMidEndSepPunct{\mcitedefaultmidpunct}
{\mcitedefaultendpunct}{\mcitedefaultseppunct}\relax
\EndOfBibitem
\bibitem[Senn and Thiel(2009)Senn, and Thiel]{senn2009qm}
Senn,~H.~M.; Thiel,~W. {QM/MM} methods for biomolecular systems. \emph{Angew.
  Chem. Int. Ed.} \textbf{2009}, \emph{48}, 1198--1229\relax
\mciteBstWouldAddEndPuncttrue
\mciteSetBstMidEndSepPunct{\mcitedefaultmidpunct}
{\mcitedefaultendpunct}{\mcitedefaultseppunct}\relax
\EndOfBibitem
\end{mcitethebibliography}
}

\end{document}